\documentclass[twocolumn]{aastex63}
\usepackage{amssymb}
\usepackage{makecell}
\usepackage{amsmath}

\received{}
\revised{}
\accepted{}

\submitjournal{ApJ}

\shorttitle{}
\shortauthors{}

\begin{document}
\title{Canonical X-Ray Fluorescence Line Intensities as Column Density Indicators}

\correspondingauthor{Roi Rahin}
\email{roir@campus.technion.ac.il}

\author{Roi Rahin}
\affil{Physics Department, Technion, Haifa 32000, Israel}

\author{Ehud Behar}
\affil{Physics Department, Technion, Haifa 32000, Israel}

\begin{abstract}
X-ray line fluorescence is ubiquitous around powerful accretion sources, namely active galactic nuclei and X-ray binaries.
The brightest and best studied line is the Fe K$\alpha$ line at $\lambda = 1.937$\AA (6.4\,keV).
This paper presents a survey of all well measured Chandra/HETG grating spectra featuring several K$\alpha$ fluorescence lines from elements between Mg and Ni.
Despite the variety of sources and physical conditions, we identify a common trend that dictates the K$\alpha$ line intensity ratios between elements.
For the most part, the line intensities are well described by a simple, plane-parallel approximation of a near-neutral, solar-abundance, high column density ($N_{\textrm{H}} > 10^{24}$ cm$^{-2}$) medium.
This approximation gives canonical photon-intensity line ratios for the K$\alpha$ fluorescence of all elements, e.g., 0.104:\,0.069:\,1.0:\,0.043 for Si:\,S:\,Fe:\,Ni, respectively.
Deviations from these ratios are shown to be primarily due to excess column density along the line of sight beyond the Galactic column. Therefore, measured fluorescence line ratios provide an independent estimate of $N_{\textrm{H}}$ and insight into the environment of accretion sources.
Residual discrepancies with the canonical ratios could be due to a variety of effects such as a fluorescing medium with $N_{\textrm{H}} < 10^{24}$\,cm$^{-2}$, a non-neutral medium, variations in the illuminating spectrum, non-solar abundances, or an irregular source geometry.
However, evidently and perhaps surprisingly, these are uncommon, and their effect remains minor.\\

\end{abstract}


\section{Introduction} \label{sec:intro}

X-ray fluorescence is a natural consequence of powerful X-ray sources illuminating ambient, cold material.
Inner-shell ionization or excitation of neutral or near-neutral ions from the K-shell may be followed by decays filling the K-shell vacancy,
and resulting in K$\alpha$ and K$\beta$ emission lines.
Near neutral ions are loosely defined here as all ions whose K$\alpha$ line can not be distinguished from that of the neutral atom with contemporary X-ray spectrometers, i.e. $\Delta \lambda \lesssim 20$ m\AA{}.
K-shell lines from near-neutral species are generally referred to as X-ray fluorescence lines.

Fluxes of K-shell fluorescence lines from various elements have been measured for a few bright sources \citep[e.g.,][]{sako1999x}, but no simple explanation exists for their relative intensities.
These intensities may depend on several parameters, such as the elemental abundances, and the fluorescence yields.
The fluorescence yield is the probability that an inner-shell excited ion will decay radiatively (i.e. fluorescence), and not auto-ionize (Auger decay), both decays being spontaneous.
For low-$Z$ elements ($Z < 12$), the fluorescence yields are small (few percent), and auto-ionization dominates.
However, the radiative decay rate scales strongly with nuclear charge ($\propto Z^4$ in the hydrogenic approximation), while auto-ionization rates are essentially independent of it \citep{bambynek1972x}.
Consequently, the fluorescence yields increase slowly with the atomic number, up to $\sim35\%$ for Fe, which produces the most prominent fluorescence K$\alpha$ line.

To complicate things, the intensities of the various fluorescence lines depend not only on the atomic properties, but also on those of the fluorescing medium. 
The X-ray intensity reaching and exciting the atoms depends on the radiative transfer of the external continuum source into this medium, which depend on the internal ionization and temperature structure of the medium \citep[][and references therein]{dopita2002narrow,stern2014radiation}.
The near-neutral species emitting fluorescence lines can survive the high X-ray field only if the ionization parameter ($L/nr^2$) is low. Here $L$ is the source luminosity, $n$ is the number density of the fluorescing medium and $r$ is its distance from the source. This requirement suggests fluorescence lines originate in the denser (possibly clumpy) parts of the medium \citep{sako1999x,sambruna2000high,sako2000chandra,yaqoob2012nature,kallman2013census,arevalo20142,xu2015x}.

The physical scale of fluorescing gas remains uncertain, even in similar-type sources, such as Seyfert 2’s. \citet{sako2000chandra} observed fluorescence from gas over hundreds of pc’s in Mrk\,3. 
\citet{sambruna2000high} found the fluorescing medium in the Circinus galaxy to be concentrated within 15\,pc of the nucleus, 
although \citet{marinucci2013chandra} and \citet{arevalo20142} later found evidence for significant flourescense by cold gas from an extended ionization cone out to $\sim$100\,pc. In NGC\,4151 \citet{ogle2000chandra} found K$\alpha$ emission extended out to $\sim$200\,pc, while \citet{miller2018x} place the emission well below 1pc.
In X-ray binaries, the geometry and morphology of the fluorescing source can be studied from the variability due to the binary orbit \citep[e.g.,][]{watanabe2006x}.

Finally, the observed fluorescence line intensity depends also on the probability of the line photons escaping the medium and into our line of sight. 
For example, resonant absorption of lines by the same atomic species that emitted them can be followed by auto-ionization, which essentially eliminates the fluorescence-line photon.
This process has been termed resonant Auger destruction, and depends on the details of the atomic state of the emitting and absorbing atoms \citep{1996MNRAS.278.1082R,liedahl2005resonant}.
The effect of radiative transfer on spectral lines is a complex problem influenced by the medium geometry and the illuminating spectrum and has been comprehensively treated by several codes \citep{george1991x,netzer1996x,kallman2001photoionization,ferland20172017}.

In this work, we aim to present all high S/N X-ray spectra of accreting sources, namely active galactic nuclei (AGN) and X-ray binaries (XRB), which feature fluorescence lines from several elements.
We focus on the astrophysically abundant elements up to Ni. 
Low-Z fluorescence from elements lighter than Mg are rarely observed, due to their low fluorescence yield. 
Hence, this paper represents a comprehensive survey of K-shell fluorescence from Mg - Ni.
The goal of the present study is to seek common attributes of fluorescence lines, despite the variety of physical conditions occurring around accretion sources, and despite the apparent complexity of the atomic states and radiative transfer. 

For this purpose, we use the archive of the Chandra/HETG grating spectrometer, for its high spectral resolution and low background.
These attributes increase the confidence in the detection of the weakest fluorescence lines. 
The targets and data reduction are described in Sec. \ref{sec:obs}.
The common themes across sources are presented in Sec. \ref{sec:Data}.
A simple plane-parallel model is described in Sec. \ref{sec:Theory}, and its adequacy for explaining the observed line intensities is presented in Sec. \ref{sec:results}.
Finally in Sec. \ref{sec:discussion} we present our conclusions.

\section{Observations} \label{sec:obs}

We use observations of the Chandra HETG with high quality spectra, which exhibit at least four significant fluorescence lines. The objects are either Seyfert 2s or high-mass XRB. All data were processed by The Chandra Grating-Data Archive and Catalog, TGCat \citep{huenemoerder2011tgcat}. In the following, we briefly introduce each target and its relevant observations. 

\subsection{The Circinus Galaxy}
The Circinus galaxy is a Seyfert 2 at redshift $z=0.001449$, which was observed by the Chandra HETG sixteen times between the years 2000 and 2009 for a total of 667268\,s. A list of observations is given in Table\,\ref{tab:obs}. 
These spectra feature both ionized and fluorescence lines and were reported by \citet{sambruna2000high} and \citet{arevalo20142}.
In a Seyfert 2 the X-ray source is an extended ionization cone and not expected to change between observations, thus we performed spectral fits simultaneously on all observed spectra.

\subsection{NGC 1068}
NGC 1068 is a Seyfert 2 ($z=0.00381$), which was observed by the Chandra HETG ten times in 2000 and in 2008 for a total of 438810\,s (see Table \ref{tab:obs}). These spectra were reported by \citet{ogle2003testing} and \citet{kallman2013census}.
We performed the data fits simultaneously on all observed spectra.

\subsection{Markarian 3}
Markarian 3 is a Seyfert 2 ($z=0.013$), which was observed by the Chandra HETG nine times in 2000 and in 2011 for a total of 389250\,s (see Table \ref{tab:obs}). These spectra were reported by \citet{sako2000chandra} and \citet{bogdan2017probing}.
We performed the data fits simultaneously on all observed spectra.

\subsection{NGC 4151}
NGC 4151 is an AGN ($z=0.003262$), which was observed by the Chandra HETG seven times. NGC\,4151 changes between obscured (Seyfert 2) and unobscured (Seyfert\,1) states. Only four HETG observations are in the obscured state, where the fluorescence lines dominate over the continuum. These spectra were reported by \citet{ogle2000chandra} and \citet{couto2016new}. Because of the high spectal variability of NGC\,4151, we analyzed each observation independently. The K$\alpha$ lines vary, especially  Fe and Si  K$\alpha$ (Table\,\ref{tab:4151_flux}), which is most likely a result of variable source luminosity. To improve readability, figures in the present paper include only a joint measurement of the last two observations, which have similar spectra.

\subsection{Vela X-1}
Vela X-1 is a high mass, eclipsing, X-ray binary. Vela X-1 was observed by the HETG eight times between 2000 and 2017. Unlike AGN, the spectrum of X-ray binaries varies significantly over short time periods \citep[e.g.,][]{belloni1990atlas,kreykenbohm2008high}. The variation is in large part due to the orbital position \citep{watanabe2006x}. As such, each observation was analyzed separately. The observations used are listed in Table \ref{tab:obs}.  
The HETG spectrum of Vela X-1 was analyzed in \citet{schulz2001ionized,goldstein2004variation,watanabe2006x}.

The HETG observed Vela X-1 in several phases. According to \citet{watanabe2006x} observation 1926 was made at phase 0.98-0.093 and 1927 at phase 0.481-0.522. Based on these data and on the Vela X-1 period of $8.964368 \pm 0.00004$ \citep{quaintrell2003mass} we estimated the phases of the other observations. We denote each observation according to its phase in Table \ref{tab:obs}

\subsection{GX 301-2}
GX 301-2 is a non-eclipsing high mass X-ray binary. GX 301-2 was observed three times by the Chandra HETG in the years 2000 and 2002. 
The GX 301-2 spectrum shows significant and variable column density \citep{mukherjee2004orbital,islam2014orbital}; however, it is hard to determine how much of it affects the fluorescence lines. The HETG observed GX 301-2 in 3 different phases: $0.167-0.179$, $0.480-0.497$, and $0.970-0.982$,  referred to by \citet{watanabe2003detection} as intermediate (IM), near-apastron (NA), and pre-periastron (PP) phases, respectively.
We include all 3 observations of GX 301-2 in the present analysis (Table \ref{tab:obs}).

\section{Data Analysis}\label{sec:Data}

We perform the spectral analysis using Xspec version 12.10.1f \citep{arnaud1996xspec}. We focus only on the total flux of the narrow fluorescence lines, which are measured by fitting a Gaussian to each line. The line widths of local Seyfert 2s in grating spectra are broadened by their angular extent, which impedes the extraction of kinematics and thus location of the fluorescing medium. However, our analysis is unaffected by kinematic or instrumental broadening. When calculating line fluxes at the source we include neutral absorption representing the Galactic column density \citep{dickey1990hi,kalberla2005leiden,bekhti2016hi4pi}, assuming solar abundances \citep{asplund2009chemical}.
The measurements are summarized in Tables \ref{tab:vela_flux}, \ref{tab:gx_agn_flux}, and \ref{tab:4151_flux}.

\subsection{Si and Mg K$\alpha$}\label{SS:SiK}
The Si K$\alpha$ line at 7.126 \AA{} blends with the Mg\,XII Ly$\beta$ line at 7.106 \AA{}. To best estimate the Si K$\alpha$ flux, we approximate the flux of the Mg\,XII Ly$\beta$ based on its Ly$\alpha$ counterpart at 8.421 \AA{}. The flux of the Ly$\beta$ line is assumed to be 20\% that of the Ly$\alpha$ line based on the Xstar code \citep{kallman2001photoionization}. We thus measure the flux in the Mg XII Ly$\alpha$ line, and subtract 20\% of it from the Si K$\alpha$ measured flux. This fraction is also supported by the analysis of \citet{liu2016clumpy}.

The Mg K$\alpha$ line at $9.890$ \AA{} suffers from low signal. In NGC\,1068 we identify another line at $\sim9.843$ \AA{} (rest frame) with comparable flux, which we include in the total Mg K$\alpha$ flux. This may indicate fluorescence from slightly ionized Mg in NGC\,1068. 

\subsection{Fe K$\beta$} \label{sec:kb}

Apart from all K$\alpha$ lines we also measure the Fe K$\beta$ fluxes. These are hampered by the diminishing effective area of the HETG and the neighboring H-like and He-like Fe lines. The Fe  K$\beta$/K$\alpha$ photon-intensity ratio ranges from 0.1-0.3 with large uncertainties (Tables \ref{tab:vela_flux} and \ref{tab:gx_agn_flux}). Laboratory measured ratios are 0.115-0.15 and theoretically up to 0.165 for Fe IX \citep[][and references therein]{palmeri2003modeling}. The present measurements are mostly consistent with these values, except for the brightest phases of GX 301-2 and Vela X-1 when the ratio is higher. This could possibly be explained with even higher ionization.

\subsection{Luminosities}

\begin{figure}[ht!]
	\includegraphics[scale=0.15]{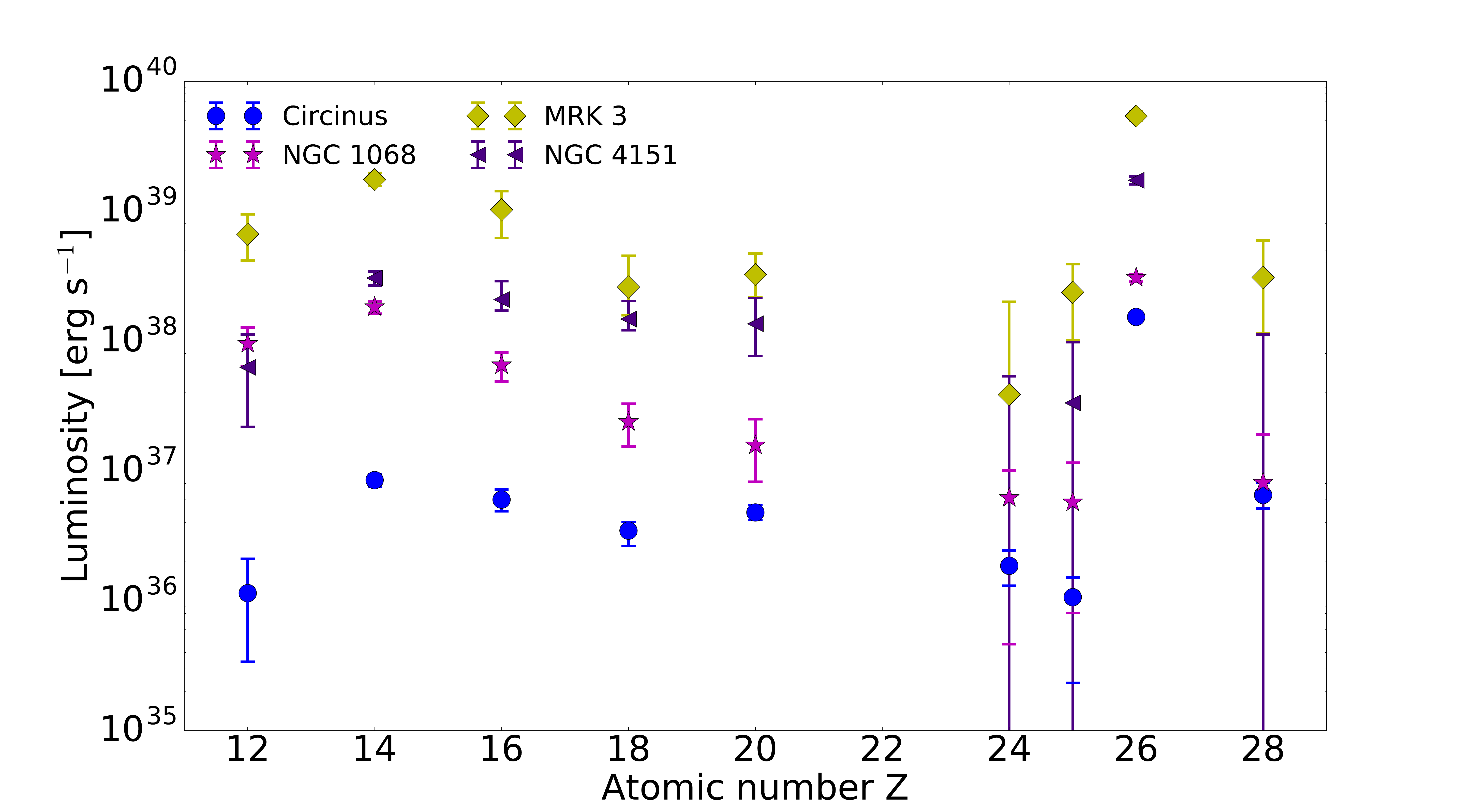}
	\includegraphics[scale=0.15]{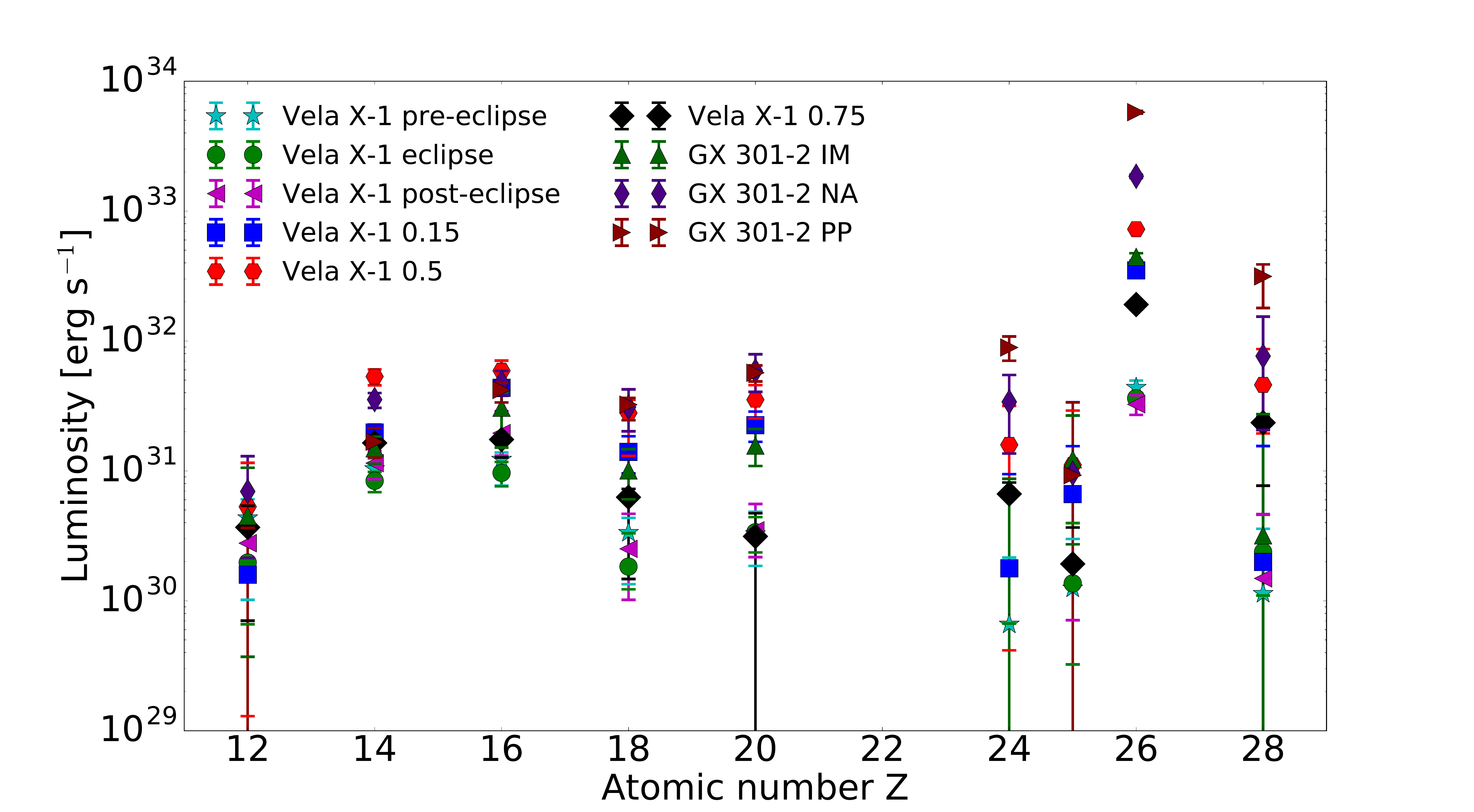}
	\caption{Luminosity of K$\alpha$ fluorescence lines corrected for Galactic absorption. \textbf{Top}: AGN. \textbf{Bottom}: XRB. Note the similar trends in AGN despite the disparity in luminosity and the phase dependence in the XRB. \label{fig:ka_lum}}
\end{figure}

We calculate the luminosity of the different fluorescence lines in various AGN and XRB by assuming isotropic emission. The results are shown in Figure\,\ref{fig:ka_lum}. Although the sources vary greatly in luminosity, we note a general decreasing intensity trend with $Z$ in the AGN and a flatter trend in the XRB. Fe K$\alpha$ remains the brightest line by far in all sources. Understanding these trends is the goal of the present paper.

\subsection{Relative intensities}

\begin{figure}[hbt!]
	\includegraphics[scale=0.15]{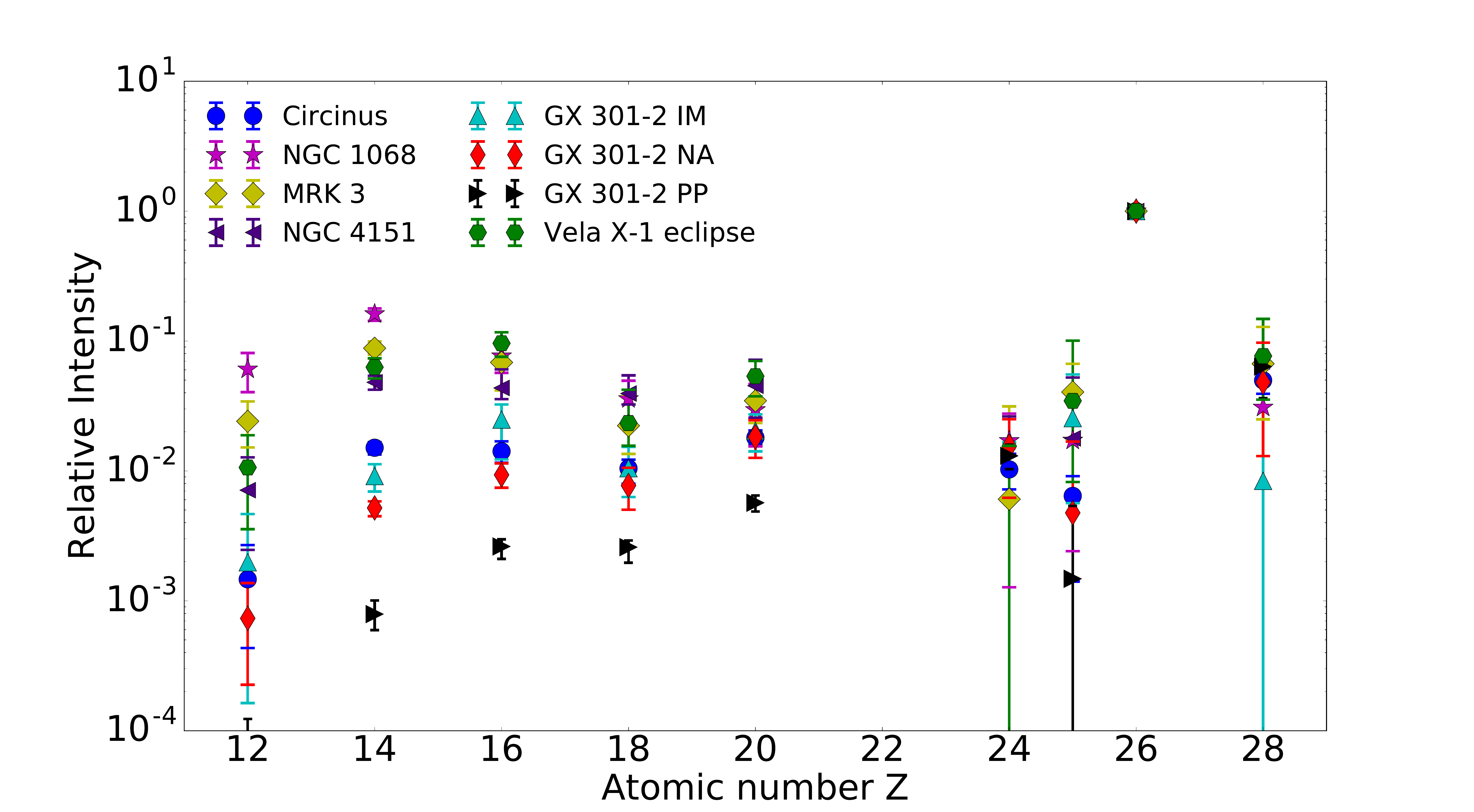}
	\includegraphics[scale=0.15]{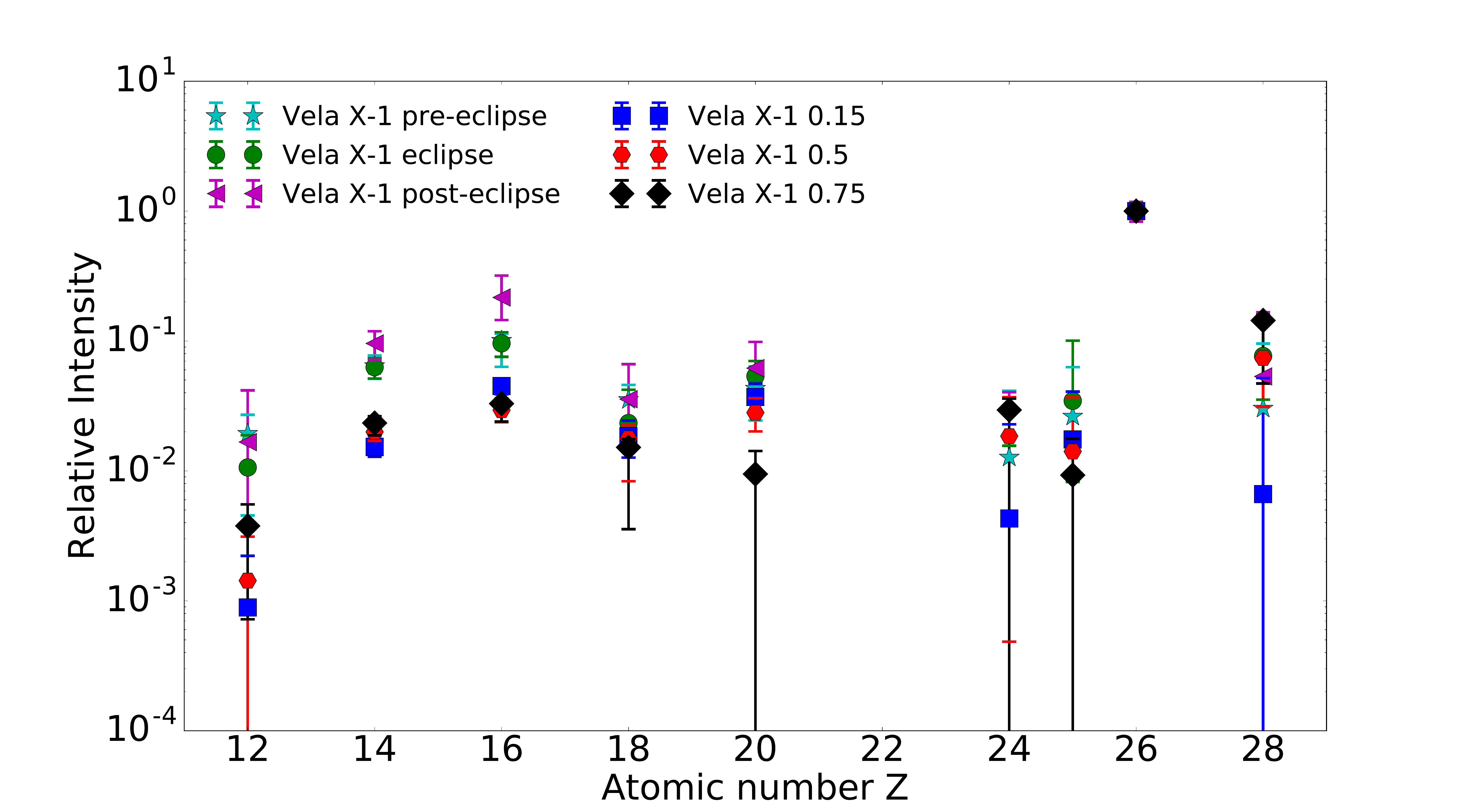}
	\caption{K$\alpha$ fluorescence line fluxes plotted relative to the flux of the Fe line. The Fe line is by far the strongest by 1 to 3 orders of magnitude depending on the source. \textbf{Top}: The source sample with a single Vela X-1 observation. The intensities do not, by themselves, reveal an inherent distinction between AGN and XRB. \textbf{Bottom}: All Vela X-1 epochs studied in this paper, suggesting a connection between line ratios and orbital phase. \label{fig:ka_flux_rel}}
\end{figure}

We aim to compare the relative strength of the fluorescence lines of various objects. We used the Fe K$\alpha$ flux in each object to normalize the fluxes, as it is the most prominent fluorescence line and thus has the smallest measurement uncertainties. The relative fluxes are shown in Figure \ref{fig:ka_flux_rel}. In the next section, we present a simple theoretical model to explain these relative intensities.

\section{Model for X-ray Fluorescence}\label{sec:Theory}

When a dense neutral medium is illuminated by an X-ray source the various elements in the medium may be excited and produce X-ray fluorescence. The fluorescence from each element in each layer in the medium is proportional to the intensity of radiation reaching that layer.

The following derivation was adapted from \citet{thomsen2007basic}. In a plane parallel approximation the incident photon-intensity $I(E,x)$ at energy $E$ and depth $x$ into the medium is:
                 
\begin{equation} 
I(E,x) = I(E,0)e^{-\tau(E,x)}
\end{equation}   
where $\tau(E,x)$ is the total optical depth a distance of $x$ into the medium, given by:

\begin{equation}
\tau(E,x) = \int_{0}^{x} \sigma(E)n_{\textrm{H}} dx = \sigma(E)N_{\textrm{H}}
\end{equation}
where $\sigma(E)$ is the abundance-weighted average photo-ionization cross section per H atom,  $n_{\textrm{H}}$ is the hydrogen number density, and $N_{\textrm{H}}$ is the hydrogen column density. 

The K-shell ionization probability of a specific neutral element $Z$ by incident radiation at depth $x$ is:
\begin{equation} 
d\tau_{Z,K} = \sigma_{Z,K}(E) n_{\textrm{Z}} dx = A_{\textrm{Z}} \sigma_{Z,K}(E) n_{\textrm{H}} dx
\end{equation}
where $\sigma_{Z,K}$ is the K-shell ionization cross section of element $Z$, $A_{\textrm{Z}}$ is the elemental abundance \cite[assumed to be solar,][]{asplund2009chemical}, and $n_{\textrm{Z}}=A_{\textrm{Z}} n_{\textrm{H}}$ is the number density of element $Z$. Thus, the intensity decrement due to K-shell ionization by an element $Z$ at a depth of $x$ in the medium is given by:
\begin{equation} 
\frac{dI}{dx} =  \int_{E_{K,Z}}^{\infty}I(E,0)e^{-\tau(E,x)}A_{\textrm{Z}}n_{\textrm{H}} \sigma_{Z,K}(E)dE
\end{equation}
where $E_{K,Z}$ is the energy of the K-edge.

After a K-shell electron is released the ion can decay mainly through K$\alpha$ fluorescence, namely a 1s-2p transition.
The probability for K$\alpha$ fluorescence  is a factor of the fluorescence yield, $\omega_K$, and the probability of a 1s-2p transition over a 1s-3p transition. Higher order transitions are relevant only for heavy elements ($Z\geq31$). This effective fluorescence yield, $\omega_{K\alpha}$, is given by:

\begin{equation} 
\omega_{K\alpha} = \omega_K\frac{I_{K\alpha}}{I_{K\alpha}+I_{K\beta}}
\end{equation}
Values of $\omega_K$ range from $\sim 0.03$ for Mg to $\sim 0.42$ for Ni. The branching ratio for K$\alpha$ over K$\beta$ emission varies from $\sim1$ for Mg to $\sim0.9$ for Ni. For $\omega_K$ we adopt the values from \citet{hubbell1994review} as they are consistent with experiments \citep{csahin2005measurement} and more recent numerical calculations \citep{palmeri2012atomic}. For the branching ratio we adopt the values of \citet{scofield1972theoretical} to ensure all values are taken from the same source. Other values \citep{PhysRevA.56.4554,csahin2005measurement,palmeri2012atomic} can change $\omega_{K\alpha}$ by $\sim2\%$ at most, which is negligible relative to the line flux uncertainties (Tables \ref{tab:vela_flux},\ref{tab:gx_agn_flux},\ref{tab:4151_flux}). As noted by \citet{palmeri2003modeling}, the branching ratio decreases with ionization. The ionization uncertainty affects $\omega_{K\alpha}$ more than that of the atomic data, but is still smaller than the measurement uncertainties.

Finally, to reach us, the emitted fluorescence photon must escape the medium. The resulting attenuation is given by the optical depth out of the medium at the energy of the K$\alpha$ line, $\tau(E_{K\alpha,Z},x)$. The observed intensity of the fluorescence line from the entire medium up to $x_{\textrm{tot}}$ is:

\begin{equation} 
\hspace*{-1cm}
I_f =  \int\displaylimits_{E_{K,Z}}^{\infty}\int\displaylimits_0^{x_{\textrm{tot}}} I(E,0)e^{-\tau(E,x)-\tau(E_{K\alpha,Z},x)}  A_{\textrm{Z}} n_{\textrm{H}} \omega_{K\alpha}\sigma_{Z,K}(E) dx  dE
\end{equation}
Solving the integral over $x$ yields:
\begin{equation} \label{eq:fluorescence}
\hspace*{-1cm}
I_f = A_{\textrm{Z}} \omega_{K\alpha} \int\displaylimits_{E_{K,Z}}^{\infty} \frac{I(E,0) \sigma_{Z,K}(E)}{\sigma(E)+\sigma(E_{K\alpha,Z})}(1-e^{-\tau(E)-\tau(E_{K\alpha,Z})})dE  
\end{equation}
where $\tau(E_{K\alpha,Z})\equiv\tau(E_{K\alpha,Z},x_{\textrm{tot}})$ is the optical depth of the medium at the K$\alpha$ line and $\tau(E)\equiv\tau(E,x_{\textrm{tot}})$ is the total optical depth of the medium. In this equation, the factors most dependent on element are $A_Z$ and $\omega_{K\alpha}$ (see Table\,\ref{tab:Canonical_Relation}). 

Two approximations for Equation \eqref{eq:fluorescence} can be used. First, we consider an optically thin medium. In this case $1-e^{-\tau(E)-\tau(E_{K\alpha,Z})} \approx \tau(E)+\tau(E_{K\alpha,Z})$. Since $\tau = \sigma  N_{\textrm H}$ Equation \eqref{eq:fluorescence} now scales linearly with $N_{\textrm H}$:
\begin{equation} 
I_f^{thin} = A_{\textrm{Z}} \omega_{K\alpha} N_{\textrm H} \int_{E_{K,Z}}^{\infty} I(E,0) \sigma_{Z,K}(E)dE
\end{equation}
We assume a power-law ionizing spectrum of the form $I(E,0) = I_0 (E/E_0)^{-\Gamma}$ and a power-law dependence of  $\sigma_{Z,K}(E) = \sigma_{Z,K}(E_{K,Z}) (E / E_{K,Z})^{-2.65}$ \citep{yaqoob2001physical} to get:

\begin{equation}   \label{eq:thin_slab}
I_f^{thin} = \frac{1}{\Gamma + 1.65} I_0 E_0^\Gamma A_{\textrm{Z}} \omega_{K\alpha} N_{\textrm H} E_{k,Z}^{1-\Gamma} \sigma_{Z,K}(E_{k,Z}) 
\end{equation}

The second approximation of Equation \eqref{eq:fluorescence} is for an optically thick medium, i.e. both $\tau(E)\rightarrow\infty$ and $\tau(E_{K\alpha,Z}) \rightarrow \infty$: 
\begin{equation} \label{eq:thick_slab}
I_f^{thick} = A_{\textrm{Z}} \omega_{K\alpha} \int\displaylimits_{E_{K,Z}}^{\infty} \frac{I_0 (\frac{E}{E_0})^{-\Gamma} \sigma_{Z,K}(E)}{\sigma(E)+\sigma(E_{K\alpha,Z})}dE
\end{equation}
This approximation turns out to be the model of reference for fluorescence lines in most astrophysical sources.  The relative K$\alpha$ line intensities predicted by Equation \eqref{eq:thick_slab} are listed in Table\, \ref{tab:Canonical_Relation}.
The values for the abundances $A_Z$ and fluorescence yields $\omega _{K\alpha}$, which drive these intensities, are also given.
It can be seen that the next strongest line after Fe K$\alpha$ is Si K$\alpha$, which is expected to be $\sim10\%$ of the intensity of the Fe line.

\begin{deluxetable}{c|c|c|c} 
	\tablecaption{Canonical $I_f^{thick}$  intensities of X-ray fluorescence lines relative to Fe K$\alpha$.}
	\label{tab:Canonical_Relation}

	\tablehead{\colhead{\textbf{Element}}&\colhead{\thead{\textbf{$\omega_{K\alpha}$}}}&\colhead{\thead{\textbf{Solar Abundance }}}&\colhead{\thead{\textbf{Relative Intensity }}}}
	\startdata
	Mg & 0.0291 & $3.98 \times 10^{-5}$ & $0.067$\\
	Si & 0.0495& $3.24 \times 10^{-5}$ &$0.104$\\
	S & 0.0768 & $1.32 \times 10^{-5}$ &$0.069$ \\
	Ar & 0.1108& $2.51 \times 10^{-6}$ &$0.021$ \\
	Ca & 0.1525& $2.19 \times 10^{-6}$ &$0.029$\\
	Ti & 0.2026& $8.91 \times 10^{-8}$ &$0.002$ \\
	Cr & 0.2587& $4.37 \times 10^{-7}$ &$0.012$\\
	Mn & 0.2870&$2.69 \times 10^{-7}$  &$0.008$\\
	Fe & 0.3164 &$3.16 \times 10^{-5}$  & $1$\\
	Ni & 0.3752 & $1.66 \times 10^{-6}$ &$0.043$\\
	\enddata
\end{deluxetable}

\begin{figure}[ht!]
	\includegraphics[scale=0.15]{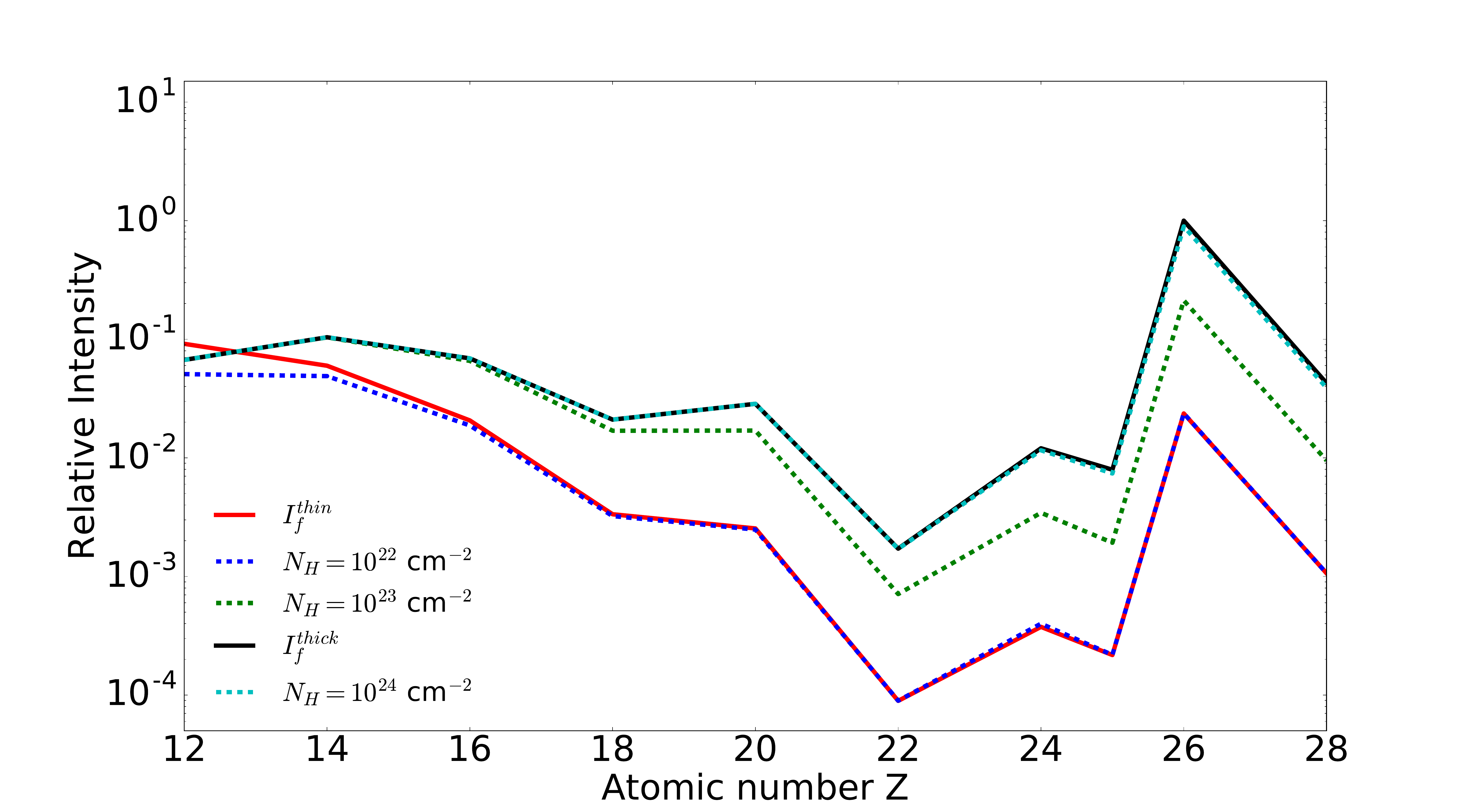}
	\includegraphics[scale=0.15]{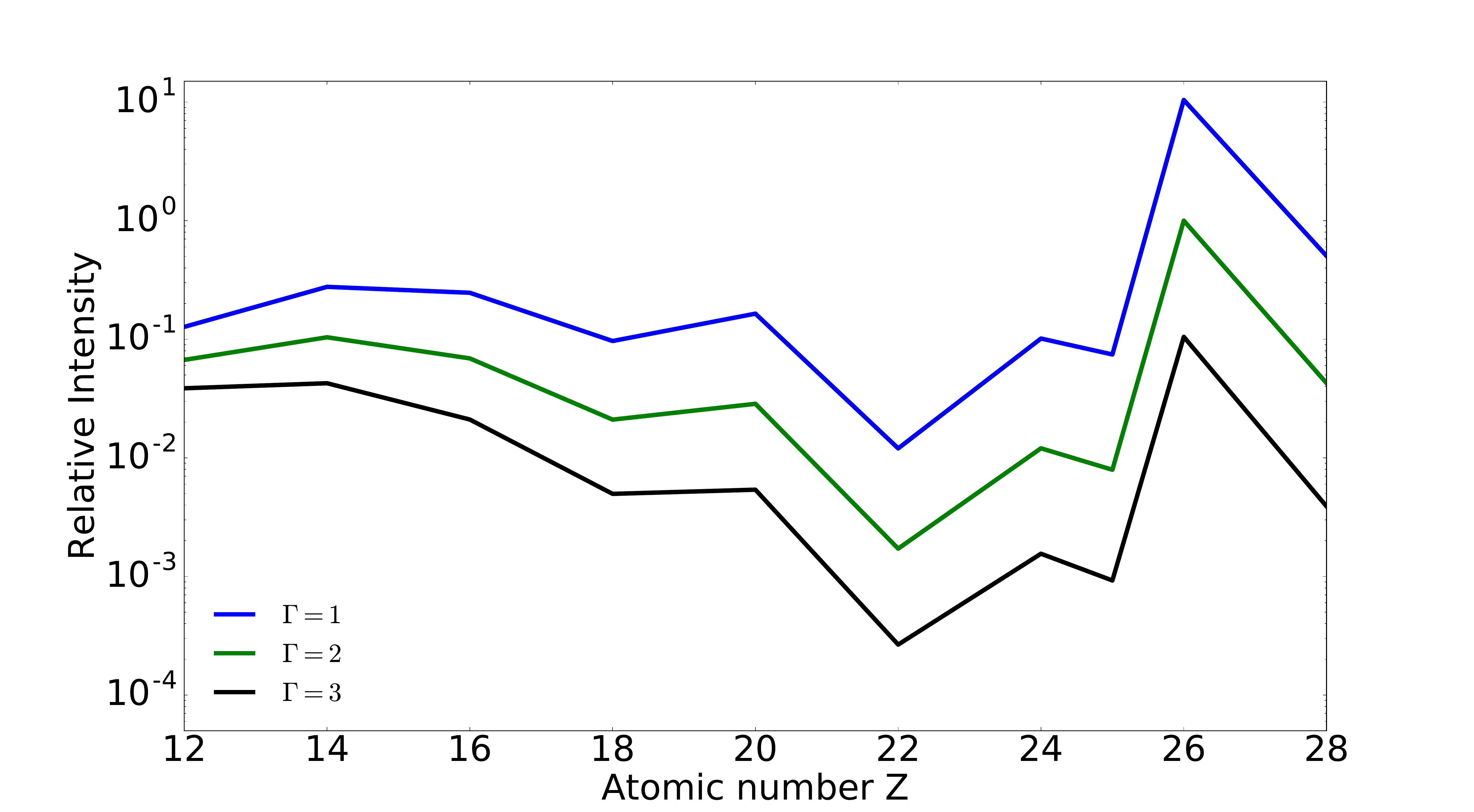}
	\caption{\textbf{Top}: Relative flux of K$\alpha$  fluorescence lines originating from different column densities of the fluorescing medium (Equation \eqref{eq:fluorescence}). All models are plotted relative to the Fe line of $I_f^{thick}$. $I_f^{thin}$ (Equation \eqref{eq:thin_slab}) agrees well with Equation \eqref{eq:fluorescence} for $N_{\textrm{H}} = 10^{22}$ cm$^{-2}$, except for Mg where $\tau(E_{K,Mg}) \ll 1$ no longer holds. $I_f^{thick}$ (Equation \eqref{eq:thick_slab}) is valid for $N_{\textrm{H}} \approx 10^{24}$ cm$^{-2}$ for all elements, and for Mg, Si, and S already at $N_{\textrm{H}} \approx 10^{23}$ cm$^{-2}$, since heavier elements emit from deeper within the medium.\\
	\textbf{Bottom}:  $I_f^{thick}$ for different power-law indices (Equation \eqref{eq:thick_slab}). All models are plotted relative to  the Fe line of the $\Gamma = 2$ model. A steeper power-law causes less ionization of high-Z elements and thus higher ratio of the low-$Z$ ($Z<26$) lines to the Fe line, and vice versa.  \label{fig:ka model}}
\end{figure}

A comparison between the approximations and the model for various column densities appear in Figure \ref{fig:ka model}, as well as a comparison between different power-law profiles.
Figure \ref{fig:ka model} shows that dramatic changes in the power law index are required to significantly change the line ratios. Note that the effect of a steep power-law is similar to that of an optically thin medium. In both cases there is a lack of hard-photon excitations.

\section{Results}  \label{sec:results}
\subsection{Comparison to theory}
\begin{figure}[ht!]
	\includegraphics[scale=0.15]{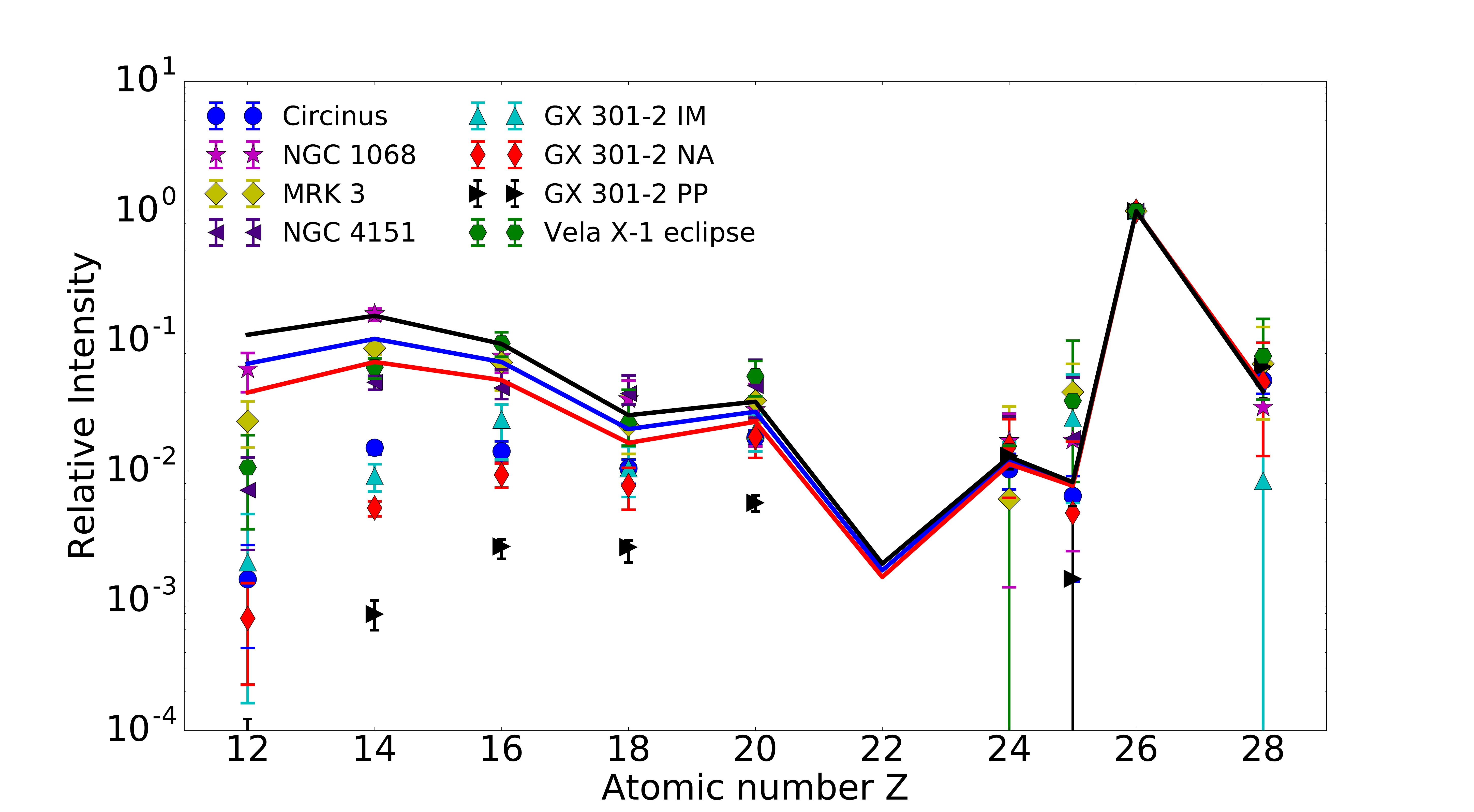}
	\includegraphics[scale=0.15]{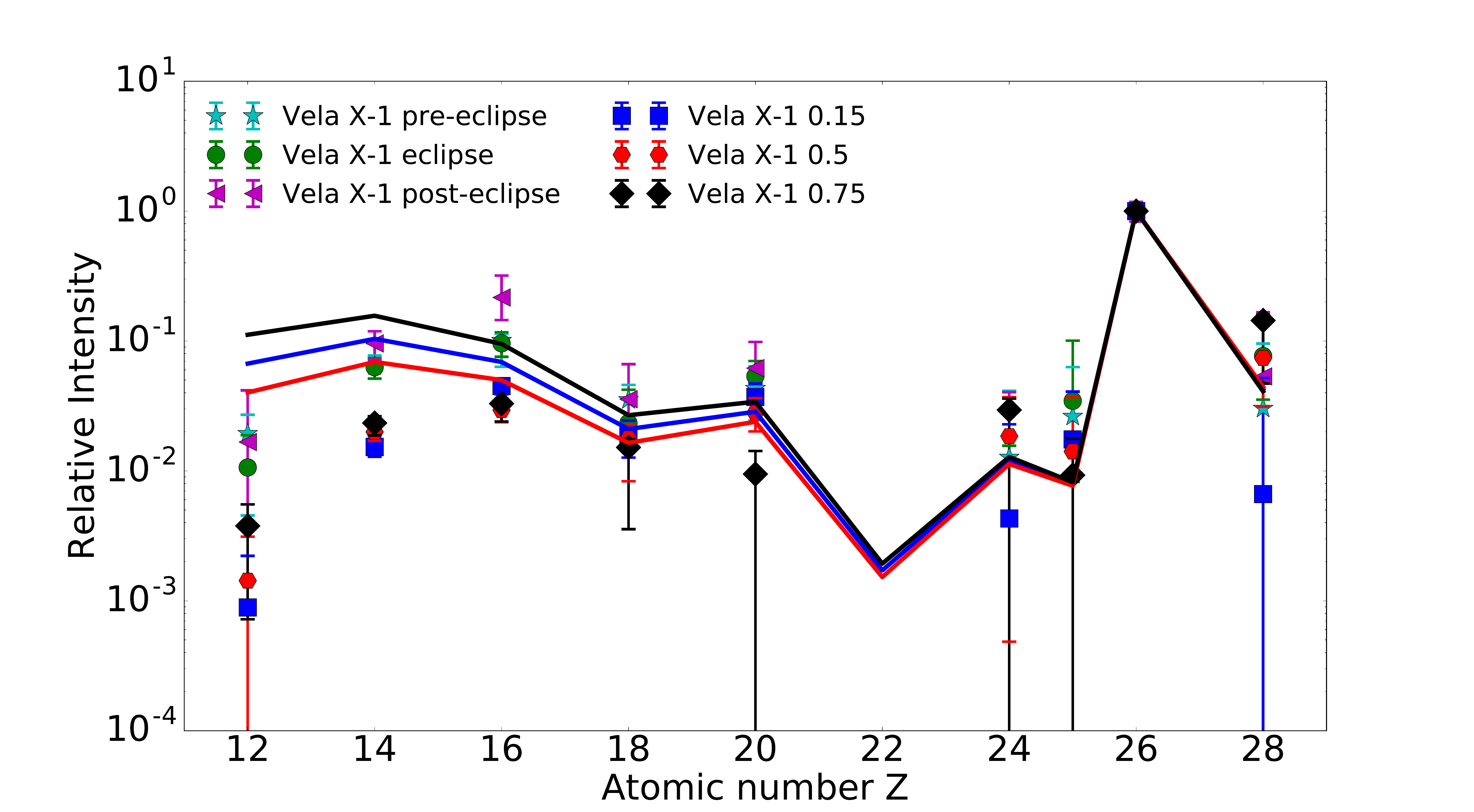}
	\caption{Relative flux of K$\alpha$ fluorescence lines (data points) compared to $I_f^{thick}$ (Equation \eqref{eq:thick_slab}) with various power-law slopes: $\Gamma = 2.3$, $\Gamma = 2 $ and $\Gamma = 1.7$ as black, blue and red lines respectively (see Figure \ref{fig:ka model}). Fluxes are plotted relative to the Fe line. \textbf{Top}: The present source sample with a single Vela X-1 epoch. The relative intensities of NGC\,1068, Mrk\,3, NGC\,4151, and Vela X-1 eclipse follow the $I_f^{thick}$ ratios, while those of GX 301-2 and Circinus show significant deficiency in the low-$Z$ lines. \textbf{Bottom}: All Vela X-1 epochs studied in this paper. Notice the deficiency in Mg, Si, and S line fluxes  in non-eclipse observations relative to $I_f^{thick}$. \label{fig:ka model compare}}
\end{figure}

The prevailing hypothesis assumes AGN to be surrounded by dense obscuring material. Some XRB, such as GX 301-2, also show significant obscuration. Such an environment calls for the approximation of an optically thick fluorescing medium. Therefore, we first compare the measured fluxes with $I_f^{thick}$ described in Equation \ref{eq:thick_slab}.
Figure \ref{fig:ka model compare} shows a comparison between the measurements and theory for three values of $\Gamma$. As can be seen, the effects of the power-law slope are moderate compared to the spread in the data.

From the graph we see that the measurements from NGC 1068,  MRK 3, and NGC 4151 are well described by Equation \eqref{eq:thick_slab}. Fluorescence lines from M51 \citep{xu2015x}, though not measured with gratings, show a similar trend. The nominal ratios imply the line-of-sight towards the fluorescing medium, as opposed to that towards the Seyfert\,2 nucleus, is unobstructed. Vela X-1 also shows a decent agreement with $I_f^{thick}$, which implies that the observed fluxes may be explained by reflection from a high $N_\textrm{H}$ wind. This is in contrast to \citet{sako1999x} who assumed  reflection from a low $N_\textrm{H}$ stellar wind, which caused the deduced $N_{\textrm{H}}$ values to increase with $Z$.

Unlike the other sources, Circinus and GX 301-2 show a systematic discrepancy of all low-Z elements with theoretical values of $I_f^{thick}$. The following sections will attempt to provide an explanation to this discrepancy.

\subsection{Resonant Auger Destruction}
\citet{sako1999x} proposed that Auger destruction may have a significant effect on fluorescence lines in Vela X-1 (see Section \ref{sec:intro}). A later theoretical analysis done by \citet{liedahl2005resonant} showed the process to be complex and highly dependent on the ionization and excitation state of each element. Indeed, K$\alpha$ lines of L-shell Si in a laboratory photo-ionized plasma experiment showed no evidence of resonant Auger destruction \citep{loisel2017benchmark}. Despite the complexity, several restrictions are immediately evident. First, K$\alpha$ Auger destruction cannot occur in a neutral medium as it requires an L-shell vacancy. Hence, higher Z elements require higher ionization for K$\alpha$ photo-excitation to occur. K$\beta$, in contrast, can be quenched by Auger destruction in lower charge states. 
Additionally, the energy difference between the K$\alpha$ energy of different L-shell Si ions is $\gtrsim 10$ eV. Thus, the velocity shear required to kinematically mix neighboring ions needs to be $\sim 1500$ km s$^{-1}$, which is not observed.

In a partially ionized medium, the ions may act as resonant absorbers for the emitted K$\alpha$ lines. Since $\omega_{K\alpha} \ll 1$, we can approximate the resulting Auger destruction as restricting our line of sight into the medium up to a photo-excitation optical depth of unity, $\tau^{PE}=1$.
The absorption cross section at the center of a resonant line under assumption of a Gaussian line profile is:

\begin{equation} 
\sigma^{PE} = \frac{\pi e^2 f}{m_e c \sqrt{2\pi} \Delta\nu} 
\end{equation}
where $e$ is the electron charge, $m_e$ is the mass of the electron, $c$ is the speed of light, $f$ is the oscillator strength and $\Delta\nu$ is the Doppler width (standard deviation). We assume $\Delta\nu/\nu$ = 0.001, which is just under the HETG resolving power.

The column density of an element $Z$ for an optical depth $\tau^{PE} =N_{\textrm{Z}}\sigma^{PE}= 1$ is:
\begin{equation} \label{eq:Nz}
N_{\textrm{Z}} = \frac{m_e c \sqrt{2\pi} \Delta\nu }{\pi e^2 f}
\end{equation}
Since $N_{\textrm{Z}}=A_{\textrm{Z}}N_{\textrm{H}}$, the corresponding $N_{\textrm H}$ for abundant elements is smaller than for rare elements. Consequently, for abundant elements $N_{\textrm H}<10^{22}$ cm$^{-2}$ and thus $\tau(E) \ll \tau^{PE} = 1$. We can therefore use Equation \eqref{eq:thin_slab} to get $I_f^{thin}$ in the case of dominant Auger destruction:

\begin{equation}  
I_f^{thin} = \frac{1}{\Gamma + 1.65} I_0 E_0^\Gamma \omega_{K\alpha} \frac{m_e c \sqrt{2\pi} \Delta\nu }{\pi e^2 f} E_{K,Z}^{1-\Gamma} \sigma_{Z,K}(E_{K,Z})\label{eq:thin_Auger}
\end{equation}
Equation \eqref{eq:thin_Auger} is independent of $A_{\textrm{Z}}$, so all abundant elements will have similar $I_f^{thin}$. 
For the rarer elements $\tau(E) \ll \tau^{PE} = 1$ no longer holds. Thus, we must use the accurate expression (Equation \ref{eq:fluorescence} with the corresponding $N_{\textrm H}$).

Another possible approximation is for Auger destruction in an optically thick medium. We assume the cross section for resonant absorption to be $\sigma_{eff,Z} = (1-\omega_{K\alpha,Z})\sigma_{Z}^{PE}$, effectively assuming the scattering is destructive with probability $(1-\omega_{K\alpha,Z})$.
This approximation effectively adds $A_Z\sigma_{eff,Z}$ to the denominator of Eq \eqref{eq:thick_slab}, which yields similar results to Equation \eqref{eq:fluorescence} with the $N_{\textrm H}$ corresponding to Equation \eqref{eq:Nz}. 

\begin{figure}[ht!]
	\includegraphics[scale=0.15]{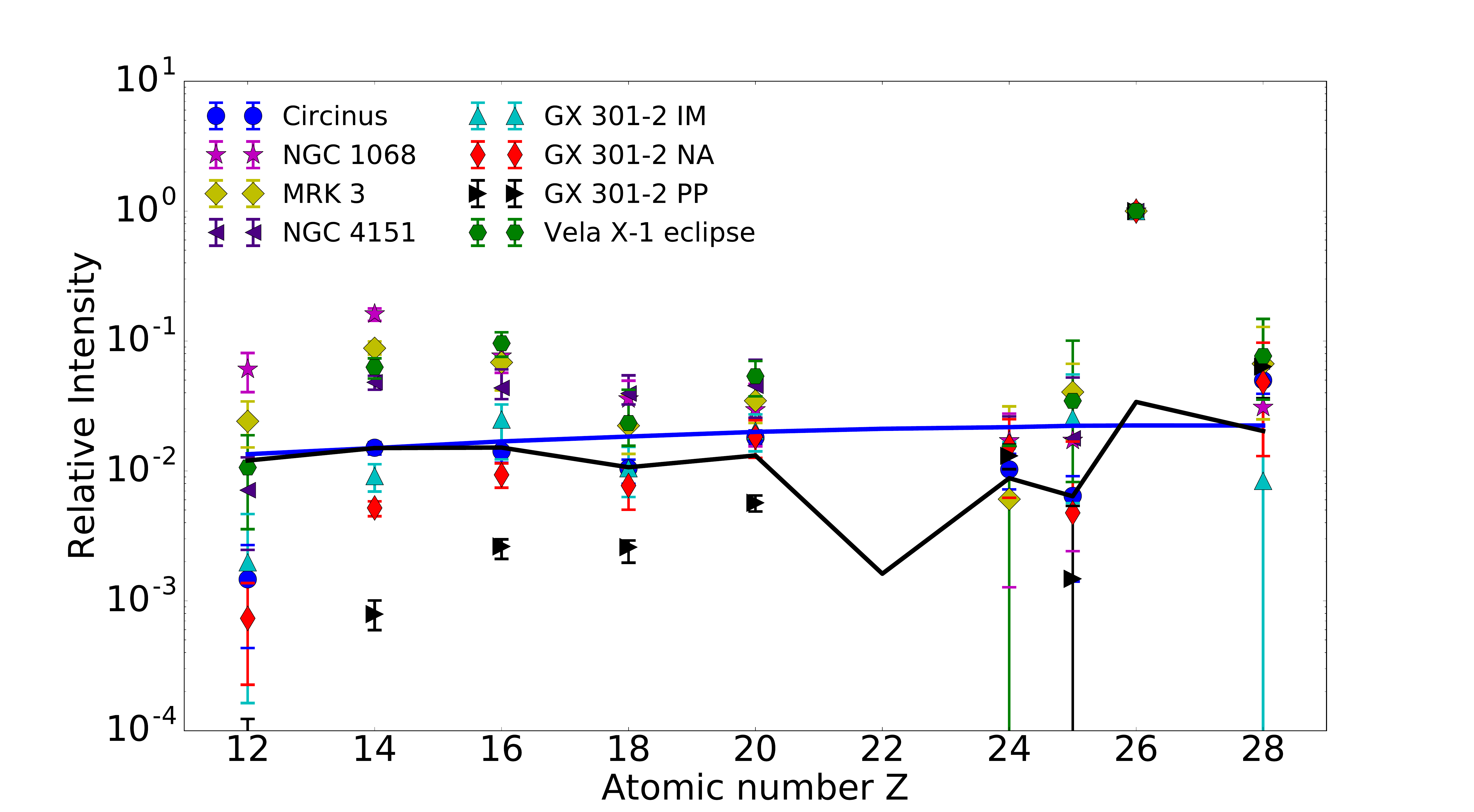}
	\includegraphics[scale=0.15]{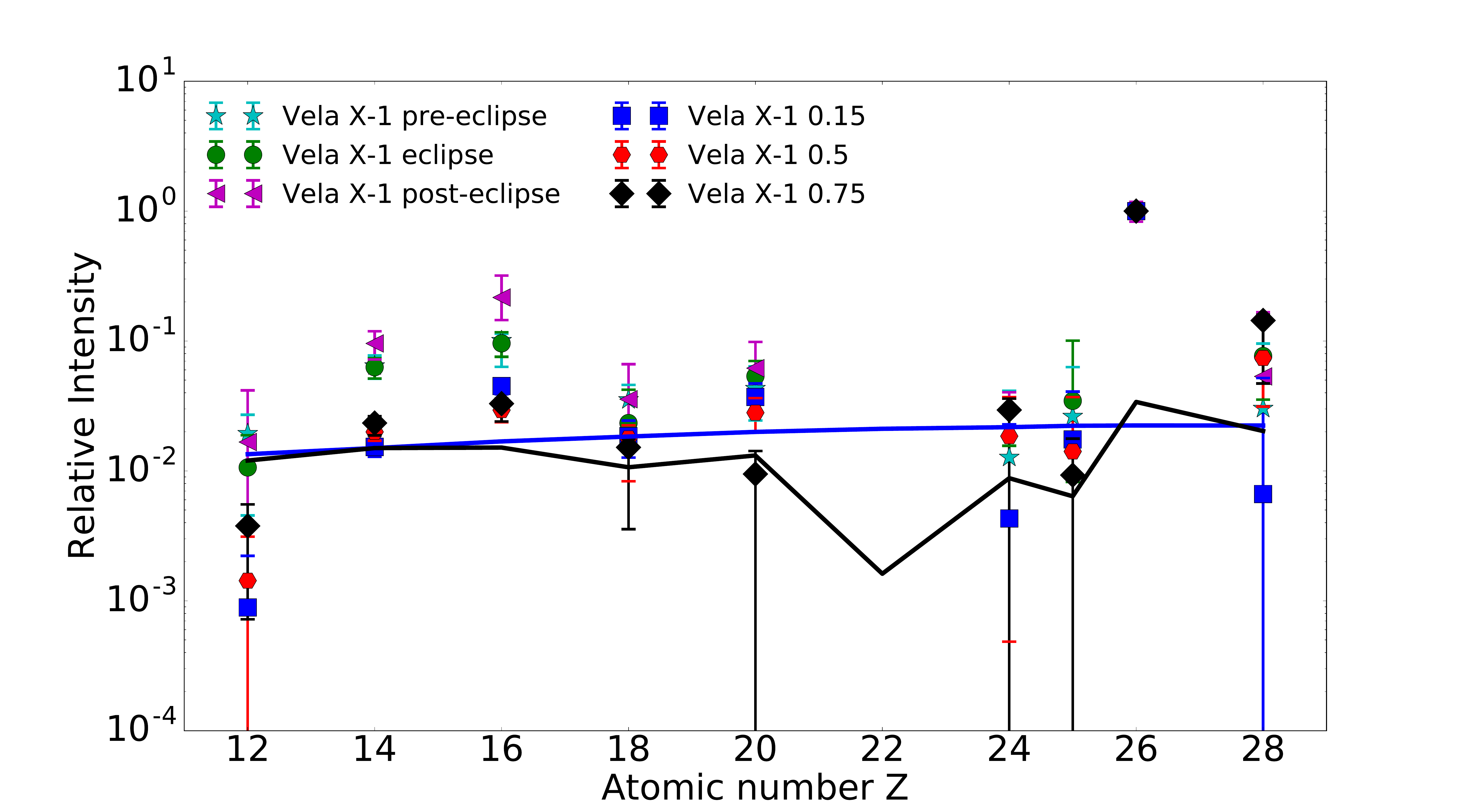}
	\caption{Relative flux of K$\alpha$  fluorescence lines compared to theoretical ratios predicted by Auger destruction models. The blue line describes $I_f^{thin}$ where the optical depth is limited by photo-excitation (Equation \eqref {eq:thin_Auger}). The black line describes  $I_f^{thick}$, but with Auger resonant scattering factored in. All models are normalized relative to the theoretical Fe line flux of $I_f^{thick}$ (Equation  \eqref {eq:thick_slab}).  \label{fig:Auger model compare}}
\end{figure}

Both approximations are shown in Figure \ref{fig:Auger model compare}, neither of which produces a good agreement with any observation. Selective Auger destruction could explain the low-Z relative intensities of Circinus, if somehow Fe is unaffected. The K$\beta$/K$\alpha$ nominal ratios presented in Sec. \ref{sec:kb} (e.g., 0.14 for Circinus) further argue against a significant role of Auger destruction, which is discussed in Section \ref{sec:discussion}. 

\subsection{Excess Line of Sight Column Density} \label{sec:losNH}
We propose a simple explanation to the relatively weak low-$Z$ lines observed in several of the present measurements. The deficiency of the low-Z relative intensities with respect to $I_f^{thick}$ (Figure \ref{fig:ka model compare}) decreases with Z. Therefore, excess line of sight column density beyond the Galactic column could explain the anomalous ratios. If correct, a single column density value $N_{\textrm H}^{ex}$ for each object would need to explain the quenching of the fluorescence lines of all low-Z elements. We can then use the measured fluxes to estimate $N_{\textrm H}^{ex}$ self-consistently for each object. The observed intensity ratio with an additional column density $N_{\textrm H}^{ex}$ would be:

\begin{equation}
\frac{I_{f,Z}^{obs}}{I_{f,Fe}^{obs}} = \frac{I_{f,Z}^{thick} e^{-\sigma(E_{K,Z}) N_{\textrm{H}}^{ex}}}{I_{f,Fe}^{thick}e^{-\sigma(E_{K,Fe}) N_{\textrm{H}}^{ex}}}
\end{equation}
Thus we can extract $N_{\textrm H}^{ex}$ for each element:

\begin{equation}\label{extra_nh_eq}
N_{\textrm H}^{ex} = \frac{1}{\sigma(E_{K,Fe})-\sigma(E_{K,Z})}\ln \left(\frac{I_{f,Z}^{obs}}{I_{f,Fe}^{obs}}\frac{I_{f,Fe}^{thick}}{I_{f,Z}^{thick}} \right)
\end{equation}
$N_{\textrm H}^{ex}$ as derived by Si K$\alpha$ is the most reliable estimate, as it is the best measured line except for Fe K$\alpha$, and their cross section difference is the second highest. For Si Equation\,\eqref{extra_nh_eq} gives:

\begin{equation}\label{eq:Si_Ex_NH}
N_{\textrm H}^{ex}(\textrm{Si}) = 2.42 \times 10^{22}\ln \left(0.104\frac{I_{f,Fe}^{obs}}{I_{f,Si}^{obs}} \right)
\end{equation}

In extra-galactic sources excess column density can be associated with dust extinction and reddening. In NGC\,1068 the above line ratios show only marginal evidence for $N_{\textrm H}^{ex}$, which is consistent with the low reddening towards the narrow line region of NGC\,1068 \citep[$A_V<0.15$,][]{crenshaw2000resolved}.
In Mrk\,3 only Mg and Si K$\alpha$ require marginal excess column density of $N_{\textrm{H}}^{ex}(\textrm{Mg})=1.1^{+0.3}_{-0.6} \times 10^{22}$ cm$^{-2}$ and $N_{\textrm{H}}^{ex}(\textrm{Si})=0.4 \pm 0.3  \times 10^{22}$ cm$^{-2}$ respectively, while other elements provide only an upper limit of $N_{\textrm{H}}^{ex}<2  \times 10^{22}$ cm$^{-2}$. This value is consistent with the reddening based estimate of \citet{collins2005physical} of $N_{\textrm H} \sim 10^{21}$ cm$^{-2}$\,.

\begin{figure}[htb!]
	\includegraphics[scale=0.15]{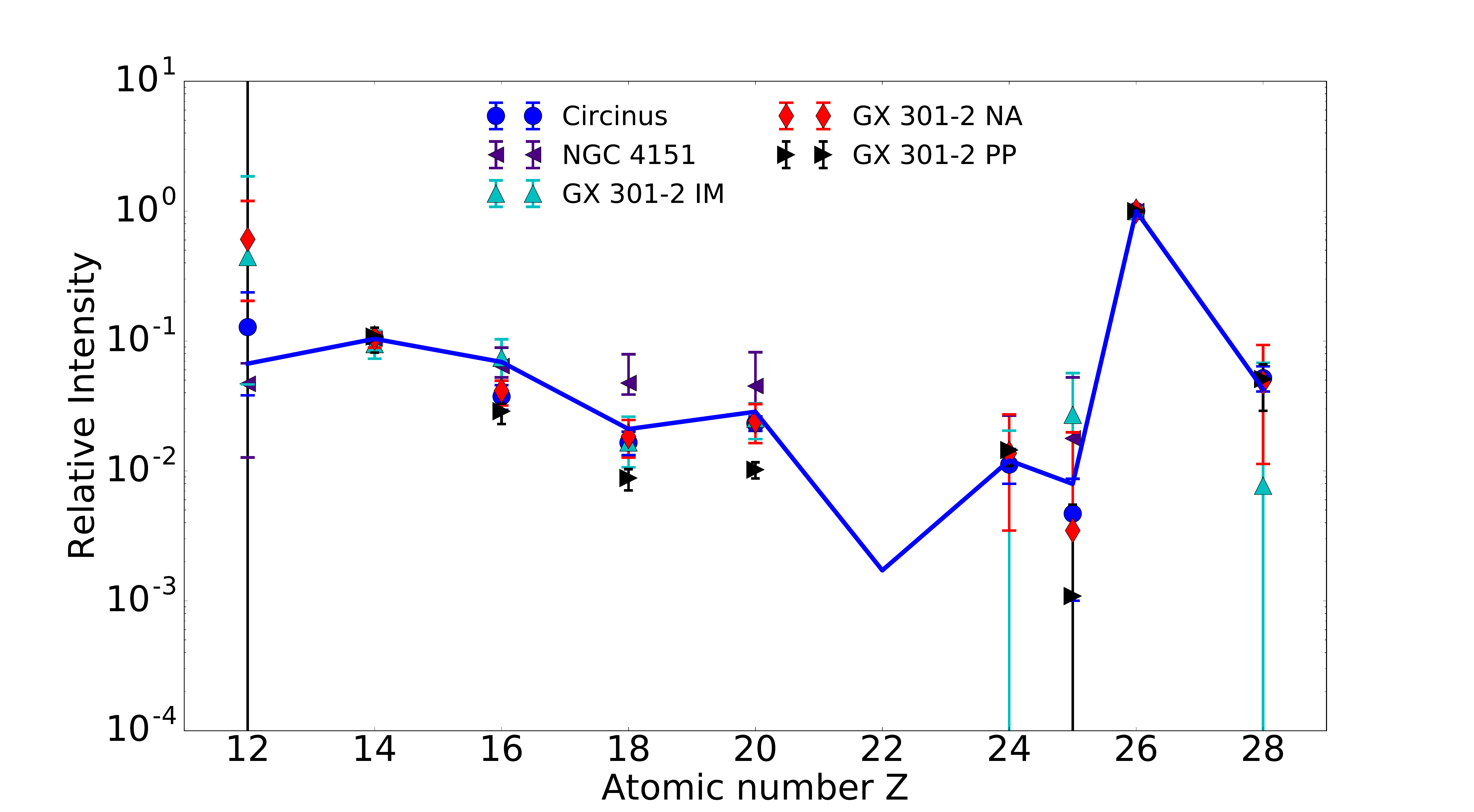}
	\includegraphics[scale=0.15]{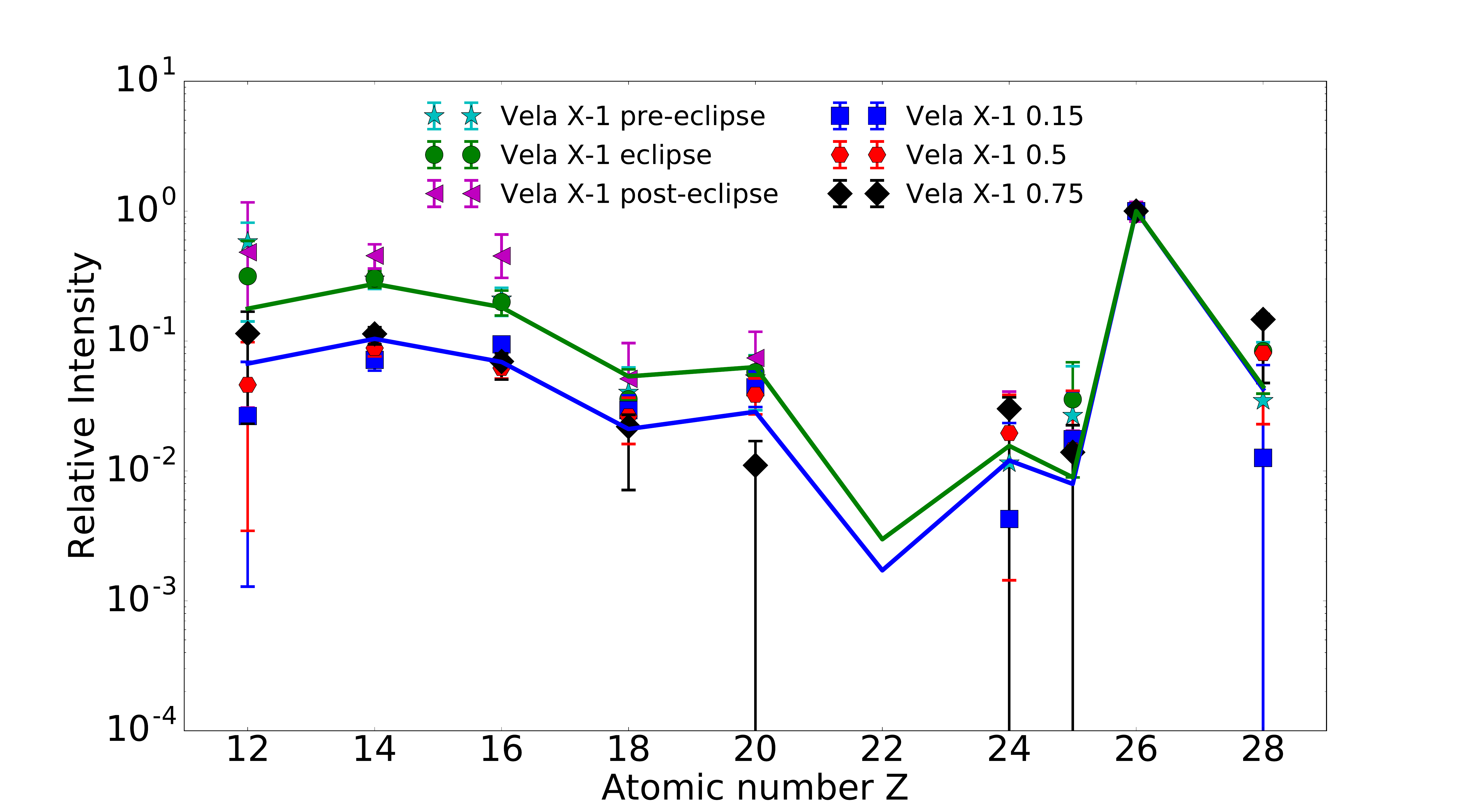}
	\caption{Relative flux of K$\alpha$ fluorescence lines compared to theoretical values after $N_{\textrm{H}}^{ex}(\textrm{Si})$ corrections. All fluxes are normalized relative to the flux of the Fe  line. The blue line represents $I_f^{thick}$ (Equation \eqref{eq:thick_slab}). \textbf{Top}:  The correction works particularly well for Circinus and the GX 301-2 IM and NA phases and less for the PP phase.
	\textbf{Bottom}: The correction works well for the Vela X-1 non-eclipse phases. For the eclipse phases a model of a fluorescing medium with $N_\textrm{H} = 2\times10^{23}$ cm$^{-2}$ is required (Equation \eqref{eq:fluorescence}, green line). \label{fig:corrected_NH}}.
\end{figure}

In the NGC\,4151 combined 2014 observations the K$\alpha$ intensity ratios of Mg, Si and S to Fe predict a column density of $N_{\textrm{H}}^{ex} \sim 2  \times 10^{22}$\,cm$^{-2}$.  The $N_{\textrm H}^{ex}$ corrected relative intensities are plotted in Figure \ref{fig:corrected_NH} (top panel). Note that Ar and Ca K$\alpha$ intensity ratios are slightly higher than canonical values and therefore require no $N_{\textrm{H}}^{ex}$.  This result is consistent with the two other observations. Since NGC\,4151 does not show significant reddening towards the narrow line region \citep{kraemer2000space}, the varying outflow may be responsible for $N_{\textrm{H}}^{ex}$. 

In contrast to the other AGN, the Circinus galaxy requires significant $N_{\textrm H}^{ex}$ (Figure \ref{fig:ka model compare}). The results are listed in Table \ref{NH_extra_AGN} and show a general agreement with each other. If we assume $\Gamma = 1.8$ instead of $\Gamma = 2.0$, the insignificant increasing trend vanishes. We find  $N_{\textrm H}^{ex}(\textrm{Si}) = (4.7 \pm 0.3)  \times 10^{22}$ cm$^{-2}$, or  $(4.0 \pm 0.3)  \times 10^{22}$ cm$^{-2}$ for $\Gamma = 1.8$, which is in excellent agreement with dust extinction measurements in Circinus. \citet{tristram2007resolving} estimated the extinction in Circinus galactic foreground to be $A_V = 20$ magnitudes which corresponds to a column density of $N_{\textrm H} \approx 4 \times 10^{22}$ cm$^{-2}$, assuming $N_{\textrm H}\approx 2A_V  \times 10^{21}$ cm$^{-2}$. The $N_{\textrm H}^{ex}$ corrected relative intensities are plotted in Figure \ref{fig:corrected_NH} (top panel) and match those predicted by $I_f^{thick}$ well. 

We note that $N_{\textrm H} > 10^{24}$ cm$^{-2}$ is the default assumption of the different AGN tori models. To fit Seyfert\,2 X-ray spectra, these models require the torus to be clumpy to create unobscured lines of sight \citep{yaqoob2012nature}. Totally unobscured cases where the Si/Fe K$\alpha$ intensity ratio is $\sim 0.1$, such as in NGC\,1068, are therefore difficult to explain under these assumptions \citep{liu2016clumpy}. The moderate $N_{\textrm H}^{ex}$ values found above for all AGN are well below the Compton-thick column of the torus. 

The situation for Vela X-1 is complicated and phase dependent. In the near eclipse observations (Eclipse, Pre and Post) deviations from $I_f^{thick}$ initially appears minor, whereas the other observations require significant $N_{\textrm H}^{ex}$ corrections. This inconsistency hints at a more complex explanation. Following \citet{sako1999x}, the fluorescing medium near eclipse is likely not optically thick and therefore not well described by $I_f^{thick}$. Thus, we must calculate excess column density based on the full model (Equation \eqref{eq:fluorescence}) rather than on $I_f^{thick}$ (Equation \eqref{eq:thick_slab}). Table \ref{NH_extra_Vela_eclipse} lists  $N_{\textrm H}^{ex}$ for the near eclipse Vela X-1 observations using a fluorescing medium of $N_{\textrm H} = 2\times10^{23}$\,cm$^{-2}$.  Table \ref{NH_extra_Vela} lists $N_{\textrm H}^{ex}$  for Vela X-1 phases 0.75, 0.5 and 0.15, calculated based on $I_f^{thick}$.  As evident in the tables, a total line of sight column density (interstellar and excess) of $N_{\textrm H}^{tot} = 4.0 \pm 0.5\times10^{22}$ cm$^{-2}$ is roughly consistent across phases and elements. In Figure \ref{fig:corrected_NH} (bottom panel) we present the relative intensities for Vela X-1 observations with $N_{\textrm H}^{tot} = 4\times10^{22}$\,cm$^{-2}$.

The most significant deviation from the $I_f^{thick}$ approximation is in GX 301-2, with each epoch requiring a different $N_{\textrm{H}}^{ex}$. The results are presented in Table\,\ref{NH_extra_GX}. Note that the Galactic $N_{\textrm{H}}$ towards GX 301-2 is already high at $N_{\textrm{H}} = 1.42\times10^{22}$\,cm$^{-2}$. The estimates for $N_{\textrm{H}}^{ex}$ for the IM and NA observations are each roughly consistent between elements. At the IM phase $N_{\textrm H}^{ex}(\textrm{Si}) = (5.9 \pm 0.6)  \times 10^{22}$ cm$^{-2}$ and at the NA phase $N_{\textrm H}^{ex}(\textrm{Si}) = (7.2 \pm 0.2)  \times 10^{22}$ cm$^{-2}$, which is consistent with MAXI measurements of total $N_{\textrm{H}} \lesssim 10^{23}$ cm$^{-2}$ and $N_{\textrm H} \approx 12 \pm 4 \times 10^{22}$ cm$^{-2}$ for the IM and NA phases respectively \citep[][Figure 5 therein]{islam2014orbital}.

The $N_{\textrm{H}}^{ex}$ values for the PP observation are higher (reaching $3\times10^{23}$\,cm$^{-2}$), as expected for an XRB near periastron and as measured by \citet{islam2014orbital}. These measurements show a trend of $N_{\textrm{H}}^{ex}$ increasing with $Z$. A possible way to remedy this inconsistency is with a flat power-law slope of $\Gamma < 1$. Indeed, $\Gamma < 1$ slopes were measured by \citet{islam2014orbital} around the PP phase.

To summarize, Figure \ref{fig:corrected_NH} shows relative intensities after correcting for $N_{\textrm{H}}^{ex}(\textrm{Si})$ in Circinus, GX 301-2 and Vela X-1. The overall agreement between the data and the models is significantly improved over that of Figure \ref{fig:ka model compare}. 

\begin{deluxetable}{c|c|c} 
	\tablecaption{Line of sight excess column density ($N_{\textrm{H}}^{ex}$, Equation \eqref{extra_nh_eq}) for the Circinus Galaxy and for the NGC\,4151 combined observation.}
	\label{NH_extra_AGN}

	\tablehead{\colhead{\textbf{Element}}&\colhead{\thead{\textbf{Circinus Galaxy} \\ \textbf{$N_{\textrm{H}}^{ex}$ $(10^{22}$ cm$^{-2})$} }}&\colhead{\thead{\textbf{NGC 4151} \\ \textbf{$N_{\textrm{H}}^{ex}$ $(10^{22}$ cm$^{-2})$} }}}
	\startdata
	Mg & $4.1\pm0.8$& $2.4 \pm 0.7$\\
	Si & $4.7 \pm 0.3$&$1.9 \pm 0.3$\\
	S & $7.9 \pm 0.9$ & $2.3\pm1.4$\\
	Ar & $7.0 \pm 2.0$& -\\
	Ca & $9.8 \pm 2.8$& -\\
	\enddata
\end{deluxetable}

\begin{deluxetable}{c|c|c|c} 
	\tablecaption{
		Line of sight excess column density ($N_{\textrm{H}}^{ex}$, Equation \eqref{extra_nh_eq}) for the Vela X-1 eclipse, pre-eclipse and post-eclipse phases, estimated based on Equation \eqref{eq:fluorescence} with $N_{\textrm H} = 2 \times 10^{23}$ cm$^{-2}$.}
	\label{NH_extra_Vela_eclipse}

	\tablehead{\colhead{\textbf{Element}}&\colhead{\thead{\textbf{Vela X-1(Eclipse)} \\ \textbf{$N_{\textrm{H}}^{ex}$ $(10^{22}$ cm$^{-2})$} }} & \colhead{\thead{\textbf{Vela X-1(Pre)} \\ \textbf{$N_{\textrm{H}}^{ex}$ $(10^{22}$ cm$^{-2})$} }}& \colhead{\thead{\textbf{Vela X-1(Post)} \\ \textbf{$N_{\textrm{H}}^{ex}$ $(10^{22}$ cm$^{-2})$} }}}
		\startdata
			Mg &  $3.0 \pm 0.8$ & $2.4 \pm 0.6$ & $2.5 \pm 1.3$ \\
			Si &  $3.6 \pm 0.4$ & $3.5 \pm 0.5$ & $2.6 \pm 0.6$ \\
			S &  $3.2 \pm 1.1$ & $2.9 \pm 1.3$ & $<1.1$\\
			Ar &  $8.3\pm5.7$ & $<8.6$ & $<11.3$ \\
			Ca &  $<10.0$ & $<17.9$ & $<10.1$\\
	\enddata
\end{deluxetable}

\begin{deluxetable}{c|c|c|c} 
	\tablecaption{Line of sight excess column density ($N_{\textrm{H}}^{ex}$, Equation \eqref{extra_nh_eq}) for Vela X-1 phases 0.5, 0.75 and 0.15, estimated based on $I_f^{thick}$ (Equation \eqref{eq:thick_slab}).}
	\label{NH_extra_Vela}
	
	\tablehead{\colhead{\textbf{Element}} & \colhead{\thead{\textbf{Vela X-1(0.5)} \\ \textbf{$N_{\textrm{H}}^{ex}$ $(10^{22}$ cm$^{-2})$} }} &\colhead{\thead{\textbf{Vela X-1(0.75)} \\ \textbf{$N_{\textrm{H}}^{ex}$ $(10^{22}$ cm$^{-2})$} }}& \colhead{\thead{\textbf{Vela X-1(0.15)} \\ \textbf{$N_{\textrm{H}}^{ex}$ $(10^{22}$ cm$^{-2})$} }} }
	\startdata
	Mg &  $4.1 \pm 1.1$ & $3.1 \pm 0.8$ & $4.6 \pm 1.3$\\
	Si &  $4 \pm 0.3$ & $3.6 \pm 0.5$ & $4.7 \pm 0.4$\\
	S &  $4.2 \pm 1$ & $3.7 \pm 1.4$ & $2.1\pm 0.7$\\
	Ar &  $<5.7$ & $<11.0 $ & $<4.4$\\
	Ca &  $<6.5$ & $-$ & $-$ \\
	\enddata
\end{deluxetable}

\begin{deluxetable}{c|c|c|c} 
	\label{NH_extra_GX}
	\tablecaption{Line of sight excess column density ($N_{\textrm{H}}^{ex}$, Equation \eqref{extra_nh_eq}) for GX 301-2.}
	
	\tablehead{\colhead{\textbf{Element}}& \colhead{\thead{\textbf{GX 301-2(IM)} \\ \textbf{$N_{\textrm{H}}^{ex}$ $(10^{22}$ cm$^{-2})$} }}  & \colhead{\thead{\textbf{GX 301-2(NA)} \\ \textbf{$N_{\textrm{H}}^{ex}$ $(10^{22}$ cm$^{-2})$} }} &\colhead{\thead{\textbf{GX 301-2 (PP)} \\ \textbf{$N_{\textrm{H}}^{ex}$ $(10^{22}$ cm$^{-2})$} }}}
	\startdata
	Mg & $3.7 \pm 1.2$ &  $4.8 \pm 0.8$ & $-$\\
	Si & $5.9 \pm 0.6$ &  $7.3 \pm 0.3$ & $11.9 \pm 0.6$\\
	S& $5.1 \pm 2$ &  $9.9 \pm 1.0$ & $16.2 \pm 0.8$ \\
	Ar & $6.9 \pm 4.3$ &  $10.0 \pm 3.6$ & $20.9 \pm 1.8$ \\
	Ca & $7.3 \pm 6.9$ &  $9.5 \pm 7.0$ & $34.5 \pm 3.0$\\
	\enddata
\end{deluxetable}

\section{Discussion} \label{sec:discussion}

Excess line of sight column density creates good agreement between the simple plane parallel models presented in Section \ref{sec:Theory} and the data across many independent targets and observations. Any minor remaining discrepancies between the data and models can be interpreted as variations in the few model assumption: ionizing spectrum, fluorescing medium $N_{\textrm{H}}$, and elemental abundances. Such examples are shown above for Vela X-1 ($N_{\textrm{H}}$) and GX 301-2 ($\Gamma$). Additionally, a more complex geometry may involve illumination-incidence and line-of-sight angles, which have a minor effect on the relative line intensities of Equation \eqref{eq:thick_slab}. An optimal fit obtained by varying all parameters is beyond the scope of the present work.

Special care should also be taken when interpreting results from the Si K$\alpha$ measurement, where we accounted for the blend with the H-like Mg Ly$\beta$ (Section \ref{SS:SiK}). In our calculations, we assumed $N_{\textrm{H}}^{ex}$ to affect only the fluorescence lines. In practice, this implies the near-neutral and ionized sources may have different line of sight column densities. This result is supported by the GX 301-2 observations, where the ionized emission lines are more obscured than the fluorescence lines. In fact, the He-like Si lines at $\sim6.7$ \AA{} are already absorbed out of the spectrum, whereas the Si K$\alpha$ is clearly visible at a longer wavelength.

Figure \ref{fig:Auger model compare} shows that Auger destruction cannot explain the observed line intensities, as it needs to selectively destroy only florescence from low-Z elements. 
Selective Auger destruction might be possible if L-shell vacancies exist in low-Z elements, but not in higher Z elements. This requirement limits the ionization parameter $\xi$ to a narrow range.  Based on \citet{kallman2001photoionization}, this range for selective Auger destruction up to and including Ca is  $1.2<\log(\xi)<1.7$, making this model restricted, but not impossible.
The main problem; however, with Auger destruction stems from the observation of the nominal Fe K$\beta$ line intensity. With selective Auger destruction, we would expect a decrease in the K$\beta$/K$\alpha$ line intensity ratio \citep{liedahl2005resonant}, which is not observed in any of the present spectra (see Section \ref{sec:kb}).

\section{Conclusions}
 The present work can be summarized as follows:

\begin{itemize}
	
	\item{We surveyed the Chandra/HETG archive for all grating spectra that feature at least four X-ray K$\alpha$ fluorescence lines in AGN and in Galactic XRB.}
	
	\item{The K$\alpha$ intensity ratios between elements follow similar trends in most observations, except for a few cases in which the low-Z lines (e.g., Mg, Si) are reduced by orders of magnitude (Figures \ref{fig:ka_lum},\ref{fig:ka_flux_rel}).}
	
	\item{For the most part, the relative K$\alpha$ intensities follow a simple plane-parallel approximation of a dense, near-neutral optically-thick medium, defined as $I_f^{thick}$ in Equation \eqref{eq:thick_slab}.
		The only two (sets of) free parameters in this model are the ionizing spectrum and the elemental abundances (Figure~\ref{fig:ka model}).}
	
	\item{The reduced K$\alpha$ intensities in the low-Z elements is explained satisfactorily and self-consistently by excess column density along the line of sight (Equation \eqref{extra_nh_eq} and Figure \ref{fig:corrected_NH}).
		This excess column derived here from the K$\alpha$ fluorescence line ratios is nicely corroborated by independent measurements, primarily reddening in AGN, and X-ray continuum absorption in XRB. Hence, X-ray fluorescence line ratios can provide an independent estimate of interstellar column density.}
	
	\item{The K$\alpha$ intensity ratios do not show evidence for resonant Auger destruction.
		Furthermore, the K$\beta$/K$\alpha$ intensity ratios are nominal, or even enhanced, in all sources.
		This further indicates the K$\alpha$ fluorescing medium is near neutral.}
	
\end{itemize}

\acknowledgments
R. R. is supported by a Ramon scholarship from the Israeli Ministry of Science and Technology.
We acknowledge support by a Center of Excellence of THE ISRAEL SCIENCE FOUNDATION (grant No. 2752/19). We thank Ari Laor, Duane Liedahl, and Richard Mushotzky for helpful comments on the manuscript, and the high school student Ahmad Ghanayim for searching and finding potential fluorescence line sources.

\vspace{5mm}

\software{Xspec \citep{arnaud1996xspec}}

\bibliographystyle{apj}
\bibliography{Bibliography.bib}

\begin{thebibliography}{}
\expandafter\ifx\csname natexlab\endcsname\relax\def\natexlab#1{#1}\fi

\bibitem[{Ar{\'e}valo {et~al.}(2014)Ar{\'e}valo, Bauer, Puccetti, Walton, Koss,
  Boggs, Brandt, Brightman, Christensen, Comastri, {et~al.}}]{arevalo20142}
Ar{\'e}valo, P., Bauer, F., Puccetti, S., {et~al.} 2014, The Astrophysical
  Journal, 791, 81

\bibitem[{Arnaud(1996)}]{arnaud1996xspec}
Arnaud, K. 1996, in Astronomical Data Analysis Software and Systems V, Vol.
  101, 17

\bibitem[{Asplund {et~al.}(2009)Asplund, Grevesse, Sauval, \&
  Scott}]{asplund2009chemical}
Asplund, M., Grevesse, N., Sauval, A.~J., \& Scott, P. 2009, Annual Review of
  Astronomy and Astrophysics, 47, 481

\bibitem[{Bambynek {et~al.}(1972)Bambynek, Crasemann, Fink, Freund, Mark,
  Swift, Price, \& Rao}]{bambynek1972x}
Bambynek, W., Crasemann, B., Fink, R., {et~al.} 1972, Reviews of modern
  physics, 44, 716

\bibitem[{Bekhti {et~al.}(2016)Bekhti, Fl{\"o}er, Keller, Kerp, Lenz, Winkel,
  Bailin, Calabretta, Dedes, Ford, {et~al.}}]{bekhti2016hi4pi}
Bekhti, N.~B., Fl{\"o}er, L., Keller, R., {et~al.} 2016, Astronomy \&
  Astrophysics, 594, A116

\bibitem[{Belloni \& Hasinger(1990)}]{belloni1990atlas}
Belloni, T., \& Hasinger, G. 1990, Astronomy and Astrophysics, 230, 103

\bibitem[{Bogd{\'a}n {et~al.}(2017)Bogd{\'a}n, Kraft, Evans, Andrade-Santos, \&
  Forman}]{bogdan2017probing}
Bogd{\'a}n, {\'A}., Kraft, R.~P., Evans, D.~A., Andrade-Santos, F., \& Forman,
  W.~R. 2017, The Astrophysical Journal, 848, 61

\bibitem[{Collins {et~al.}(2005)Collins, Kraemer, Crenshaw, Ruiz, Deo, \&
  Bruhweiler}]{collins2005physical}
Collins, N.~R., Kraemer, S.~B., Crenshaw, D.~M., {et~al.} 2005, The
  Astrophysical Journal, 619, 116

\bibitem[{Couto {et~al.}(2016)Couto, Kraemer, Turner, \&
  Crenshaw}]{couto2016new}
Couto, J., Kraemer, S., Turner, T., \& Crenshaw, D. 2016, The Astrophysical
  Journal, 833, 191

\bibitem[{Crenshaw \& Kraemer(2000)}]{crenshaw2000resolved}
Crenshaw, D.~M., \& Kraemer, S.~B. 2000, The Astrophysical Journal, 532, 247

\bibitem[{Dickey \& Lockman(1990)}]{dickey1990hi}
Dickey, J.~M., \& Lockman, F.~J. 1990, Annual review of astronomy and
  astrophysics, 28, 215

\bibitem[{Dopita {et~al.}(2002)Dopita, Groves, Sutherland, Binette, \&
  Cecil}]{dopita2002narrow}
Dopita, M.~A., Groves, B.~A., Sutherland, R.~S., Binette, L., \& Cecil, G.
  2002, The Astrophysical Journal, 572, 753

\bibitem[{Ferland {et~al.}(2017)Ferland, Chatzikos, Guzm{\'a}n, Lykins,
  Van~Hoof, Williams, Abel, Badnell, Keenan, Porter,
  {et~al.}}]{ferland20172017}
Ferland, G., Chatzikos, M., Guzm{\'a}n, F., {et~al.} 2017, Revista mexicana de
  astronom{\'\i}a y astrof{\'\i}sica, 53

\bibitem[{George \& Fabian(1991)}]{george1991x}
George, I., \& Fabian, A. 1991, Monthly Notices of the Royal Astronomical
  Society, 249, 352

\bibitem[{Goldstein {et~al.}(2004)Goldstein, Huenemoerder, \&
  Blank}]{goldstein2004variation}
Goldstein, G., Huenemoerder, D.~P., \& Blank, D. 2004, The Astronomical
  Journal, 127, 2310

\bibitem[{H\"olzer {et~al.}(1997)H\"olzer, Fritsch, Deutsch, H\"artwig, \&
  F\"orster}]{PhysRevA.56.4554}
H\"olzer, G., Fritsch, M., Deutsch, M., H\"artwig, J., \& F\"orster, E. 1997,
  Phys. Rev. A, 56, 4554

\bibitem[{Hubbell {et~al.}(1994)Hubbell, Trehan, Singh, Chand, Mehta, Garg,
  Garg, Singh, \& Puri}]{hubbell1994review}
Hubbell, J., Trehan, P., Singh, N., {et~al.} 1994, Journal of Physical and
  Chemical Reference Data, 23, 339

\bibitem[{Huenemoerder {et~al.}(2011)Huenemoerder, Mitschang, Dewey, Nowak,
  Schulz, Nichols, Davis, Houck, Marshall, Noble,
  {et~al.}}]{huenemoerder2011tgcat}
Huenemoerder, D.~P., Mitschang, A., Dewey, D., {et~al.} 2011, The Astronomical
  Journal, 141, 129

\bibitem[{Islam \& Paul(2014)}]{islam2014orbital}
Islam, N., \& Paul, B. 2014, Monthly Notices of the Royal Astronomical Society,
  441, 2539

\bibitem[{Kalberla {et~al.}(2005)Kalberla, Burton, Hartmann, Arnal, Bajaja,
  Morras, \& P{\"o}ppel}]{kalberla2005leiden}
Kalberla, P.~M., Burton, W., Hartmann, D., {et~al.} 2005, Astronomy \&
  Astrophysics, 440, 775

\bibitem[{Kallman \& Bautista(2001)}]{kallman2001photoionization}
Kallman, T., \& Bautista, M. 2001, The Astrophysical Journal Supplement Series,
  133, 221

\bibitem[{Kallman {et~al.}(2013)Kallman, Evans, Marshall, Canizares,
  Longinotti, Nowak, \& Schulz}]{kallman2013census}
Kallman, T., Evans, D.~A., Marshall, H., {et~al.} 2013, The Astrophysical
  Journal, 780, 121

\bibitem[{Kraemer {et~al.}(2000)Kraemer, Crenshaw, Hutchings, Gull, Kaiser,
  Nelson, \& Weistrop}]{kraemer2000space}
Kraemer, S., Crenshaw, D., Hutchings, J., {et~al.} 2000, The Astrophysical
  Journal, 531, 278

\bibitem[{Kreykenbohm {et~al.}(2008)Kreykenbohm, Wilms, Kretschmar,
  Torrej{\'o}n, Pottschmidt, Hanke, Santangelo, Ferrigno, \&
  Staubert}]{kreykenbohm2008high}
Kreykenbohm, I., Wilms, J., Kretschmar, P., {et~al.} 2008, Astronomy \&
  Astrophysics, 492, 511

\bibitem[{Liedahl(2005)}]{liedahl2005resonant}
Liedahl, D.~A. 2005in , American Institute of Physics, 99--108

\bibitem[{Liu {et~al.}(2016)Liu, Liu, Li, Xu, Gou, \& Cheng}]{liu2016clumpy}
Liu, J., Liu, Y., Li, X., {et~al.} 2016, Monthly Notices of the Royal
  Astronomical Society: Letters, 459, L100

\bibitem[{Loisel {et~al.}(2017)Loisel, Bailey, Liedahl, Fontes, Kallman,
  Nagayama, Hansen, Rochau, Mancini, \& Lee}]{loisel2017benchmark}
Loisel, G.~P., Bailey, J.~E., Liedahl, D., {et~al.} 2017, Physical Review
  Letters, 119, 075001

\bibitem[{Marinucci {et~al.}(2013)Marinucci, Miniutti, Bianchi, Matt, \&
  Risaliti}]{marinucci2013chandra}
Marinucci, A., Miniutti, G., Bianchi, S., Matt, G., \& Risaliti, G. 2013,
  Monthly Notices of the Royal Astronomical Society, 436, 2500

\bibitem[{Miller {et~al.}(2018)Miller, Cackett, Zoghbi, Barret, Behar,
  Brenneman, Fabian, Kaastra, Lohfink, Mushotzky, {et~al.}}]{miller2018x}
Miller, J., Cackett, E., Zoghbi, A., {et~al.} 2018, The Astrophysical Journal,
  865, 97

\bibitem[{Mukherjee \& Paul(2004)}]{mukherjee2004orbital}
Mukherjee, U., \& Paul, B. 2004, Astronomy \& Astrophysics, 427, 567

\bibitem[{Netzer(1996)}]{netzer1996x}
Netzer, H. 1996, The Astrophysical Journal, 473, 781

\bibitem[{Ogle {et~al.}(2003)Ogle, Brookings, Canizares, Lee, \&
  Marshall}]{ogle2003testing}
Ogle, P., Brookings, T., Canizares, C., Lee, J., \& Marshall, H. 2003,
  Astronomy \& Astrophysics, 402, 849

\bibitem[{Ogle {et~al.}(2000)Ogle, Marshall, Lee, \&
  Canizares}]{ogle2000chandra}
Ogle, P., Marshall, H., Lee, J., \& Canizares, C. 2000, The Astrophysical
  Journal Letters, 545, L81

\bibitem[{Palmeri {et~al.}(2003)Palmeri, Mendoza, Kallman, Bautista, \&
  Mel{\'e}ndez}]{palmeri2003modeling}
Palmeri, P., Mendoza, C., Kallman, T., Bautista, M., \& Mel{\'e}ndez, M. 2003,
  Astronomy \& Astrophysics, 410, 359

\bibitem[{Palmeri {et~al.}(2012)Palmeri, Quinet, Mendoza, Bautista,
  Garc{\'\i}a, Witthoeft, \& Kallman}]{palmeri2012atomic}
Palmeri, P., Quinet, P., Mendoza, C., {et~al.} 2012, Astronomy \& Astrophysics,
  543, A44

\bibitem[{Quaintrell {et~al.}(2003)Quaintrell, Norton, Ash, Roche, Willems,
  Bedding, Baldry, \& Fender}]{quaintrell2003mass}
Quaintrell, H., Norton, A.~J., Ash, T., {et~al.} 2003, Astronomy \&
  Astrophysics, 401, 313

\bibitem[{{Ross} {et~al.}(1996){Ross}, {Fabian}, \&
  {Brandt}}]{1996MNRAS.278.1082R}
{Ross}, R.~R., {Fabian}, A.~C., \& {Brandt}, W.~N. 1996, \mnras, 278, 1082

\bibitem[{{\c{S}}ahin {et~al.}(2005){\c{S}}ahin, Demir, \&
  Budak}]{csahin2005measurement}
{\c{S}}ahin, M., Demir, L., \& Budak, G. 2005, Applied radiation and isotopes,
  63, 141

\bibitem[{Sako {et~al.}(2000)Sako, Kahn, Paerels, \& Liedahl}]{sako2000chandra}
Sako, M., Kahn, S.~M., Paerels, F., \& Liedahl, D.~A. 2000, The Astrophysical
  Journal Letters, 543, L115

\bibitem[{Sako {et~al.}(1999)Sako, Liedahl, Kahn, \& Paerels}]{sako1999x}
Sako, M., Liedahl, D.~A., Kahn, S.~M., \& Paerels, F. 1999, The Astrophysical
  Journal, 525, 921

\bibitem[{Sambruna {et~al.}(2000)Sambruna, Netzer, Kaspi, Brandt, Chartas,
  Garmire, Nousek, \& Weaver}]{sambruna2000high}
Sambruna, R.~M., Netzer, H., Kaspi, S., {et~al.} 2000, The Astrophysical
  Journal Letters, 546, L13

\bibitem[{Schulz {et~al.}(2001)Schulz, Canizares, Lee, \&
  Sako}]{schulz2001ionized}
Schulz, N.~S., Canizares, C.~R., Lee, J.~C., \& Sako, M. 2001, The
  Astrophysical Journal Letters, 564, L21

\bibitem[{Scofield(1972)}]{scofield1972theoretical}
Scofield, J.~H. 1972, THEORETICAL RADIATIVE TRANSITION RATES FOR K-AND L-SHELL
  X RAYS., Tech. rep., California Univ., Livermore. Lawrence Livermore Lab.

\bibitem[{Stern {et~al.}(2014)Stern, Laor, \& Baskin}]{stern2014radiation}
Stern, J., Laor, A., \& Baskin, A. 2014, Monthly Notices of the Royal
  Astronomical Society, 438, 901

\bibitem[{Thomsen(2007)}]{thomsen2007basic}
Thomsen, V. 2007, Spectroscopy, 22, 46

\bibitem[{Tristram {et~al.}(2007)Tristram, Meisenheimer, Jaffe, Schartmann,
  Rix, Leinert, Morel, Wittkowski, R{\"o}ttgering, Perrin,
  {et~al.}}]{tristram2007resolving}
Tristram, K., Meisenheimer, K., Jaffe, W., {et~al.} 2007, Astronomy \&
  Astrophysics, 474, 837

\bibitem[{Watanabe {et~al.}(2003)Watanabe, Sako, Ishida, Ishisaki, Kahn,
  Kohmura, Morita, Nagase, Paerels, \& Takahashi}]{watanabe2003detection}
Watanabe, S., Sako, M., Ishida, M., {et~al.} 2003, The Astrophysical Journal
  Letters, 597, L37

\bibitem[{Watanabe {et~al.}(2006)Watanabe, Sako, Ishida, Ishisaki, Kahn,
  Kohmura, Nagase, Paerels, \& Takahashi}]{watanabe2006x}
---. 2006, The Astrophysical Journal, 651, 421

\bibitem[{Xu {et~al.}(2015)Xu, Liu, Gou, \& Liu}]{xu2015x}
Xu, W., Liu, Z., Gou, L., \& Liu, J. 2015, Monthly Notices of the Royal
  Astronomical Society: Letters, 455, L26

\bibitem[{Yaqoob(2012)}]{yaqoob2012nature}
Yaqoob, T. 2012, Monthly Notices of the Royal Astronomical Society, 423, 3360

\bibitem[{Yaqoob {et~al.}(2001)Yaqoob, George, Nandra, Turner, Serlemitsos, \&
  Mushotzky}]{yaqoob2001physical}
Yaqoob, T., George, I., Nandra, K., {et~al.} 2001, The Astrophysical Journal,
  546, 759

\end{thebibliography}

\appendix
\section{Data and measurements}

\begin{deluxetable*}{c|c|c|c|c}
	\tablecaption{List of observations used}
	\label{tab:obs}

	\tablehead{\colhead{\textbf{Object}} & \colhead{\textbf{Observation ID}} &\colhead{\textbf{Date observed}} & \colhead{\textbf{Exposure (s)}} & \colhead{\textbf{Notes}}}
	\startdata
	Circinus Galaxy & 374 & 2000-06-15 & 7118&\\
	Circinus Galaxy & 62877  & 2000-06-16& 60220& \\
	Circinus Galaxy & 4770 & 2004-06-02 & 55030 & \\
	Circinus Galaxy & 4771 & 2004-11-28 & 58970 & \\
	Circinus Galaxy & 10226 & 2008-12-08 & 19670 & \\
	Circinus Galaxy & 10223 & 2008-12-15 & 102900 & \\
	Circinus Galaxy & 10832 & 2008-12-19 & 20610 & \\
	Circinus Galaxy & 10833 & 2008-12-22 & 28360 & \\
	Circinus Galaxy & 10224 & 2008-12-23 & 77100 & \\
	Circinus Galaxy & 10844 & 2008-12-24 & 27170 & \\
	Circinus Galaxy & 10225 & 2008-12-26 & 67890 & \\
	Circinus Galaxy & 10842 & 2008-12-27 & 36740 & \\
	Circinus Galaxy & 10843 & 2008-12-29 & 57010 & \\
	Circinus Galaxy & 10873 & 2009-03-01 & 18100 & \\
	Circinus Galaxy & 10850 & 2009-03-03 & 13850 & \\
	Circinus Galaxy & 10872 & 2009-03-04 & 16530 & \\
	NGC 1068 & 332 & 2000-12-04 & 45700 & \\
	NGC 1068 & 9148 & 2008-12-05 & 79540 & \\
	NGC 1068 & 9149 & 2008-11-19 & 88730 & \\
	NGC 1068 & 9150 & 2008-11-27 & 41080 & \\
	NGC 1068 & 10815 & 2008-11-20 & 19070 & \\
	NGC 1068 & 10816 & 2008-11-18 & 16160 & \\
	NGC 1068 & 10817 & 2008-11-22 & 32650 & \\
	NGC 1068 & 10823 & 2008-11-25 & 34540 & \\
	NGC 1068 & 10829 & 2008-11-30 & 38440 & \\
	NGC 1068 & 10830 & 2008-12-03 & 42900 & \\
	MRK 3 & 873 & 2000-03-18 & 100600 & \\
	MRK 3 & 12874 & 2011-04-19 & 77060 & \\
	MRK 3 & 12875 & 2011-04-25 & 29860 & \\
	MRK 3 & 13264 & 201-04-27 & 35760 & \\
	MRK 3 & 13263 & 2011-04-28 & 19720 & \\
	MRK 3 & 13261 & 2011-05-02 & 22080 & \\
	MRK 3 & 13406 & 2011-05-03 & 21430 & \\
	MRK 3 & 13254 & 2011-08-26 & 31530 & \\
	MRK 3 & 14331 & 2011-08-28 & 51210 & \\
	NGC 4151 & 335 & 2000-03-05& 47440 & \\
	NGC 4151 & 7829 & 2007-03-19& 49230 & \\
	NGC 4151 & 16089 & 2014-02-12& 171900 & \\
	NGC 4151 & 16090 & 2014-03-08& 68870 & \\
	GX 301-2 & 103 & 2000-06-19 & 39519 & 0.167-179 (IM)\\
	GX 301-2 & 2733 & 2002-01-13 & 39230 & 0.97-0.982 (PP)\\
	GX 301-2 & 3433 & 2002-02-03 & 59030 & 0.48-0.497 (NA) \\
	Vela X-1 & 102 & 2000-04-13 & 28010 & 0.015-0.051 : post-eclipse (Post)\\
	Vela X-1 & 1926 & 2001-02-11 & 83150 & 0.980 - 0.093 : eclipse \\
	Vela X-1 & 1927 & 2001-02-07 & 29430 & 0.481-0.522 : 0.5  \\
	Vela X-1 & 14654 & 2013-07-30 & 45880 & 0.748-0.807 : 0.75 \\
	Vela X-1 & 18617 & 2017-02-21 & 44180 & 0.936-0.993 : pre-eclipse (Pre) \\
	Vela X-1 & 19953 & 2017-02-22 & 70470 & 0.132-0.223 : 0.15 \\
	\enddata
	
\end{deluxetable*}

\begin{deluxetable*}{c|c|c|c|c|c|c|c}
	\label{tab:vela_flux}
	\tablecaption{Vela X-1 measured fluorescence line fluxes at various orbital phases}
	\tablehead{\colhead{\textbf{Element}} & \colhead{\textbf{Rest Frame}} &\multicolumn{5}{c}{\textbf{Photon Flux ($10 ^{-6}$ ph s$^{-1}$ cm $^{-2}$)}}\\ \colhead{}& \textbf{Wavelength (\AA)} &\textbf{Eclipse}&\textbf{Pre-Eclipse}&\textbf{Post-Eclipse}&\textbf{0.5}&\textbf{0.75}&\textbf{0.15}}
	\startdata
		Mg &  9.890 & $1.9^{+1.5}_{-1.3}$& $4.2^{+1.6}_{-3.2}$& $<6.7$& $5.1^{+6.0}_{-5.0}$&$3.5^{+2.3}_{-2.9}$ & $<3.8$\\
	    Si & 7.126 & $11.2^{+1.9}_{-2.0}$ &$13.9^{+2.8}_{-2.8}$ &$15.3^{+3.7}_{-3.8}$ & $71^{+10}_{-10}$& $21.9^{+4.3}_{-4.3}$& $26.3^{+4.1}_{-4.2}$\\
		S & 5.373 & $17.1^{+3.7}_{-3.6}$  &$21.7^{+2.8}_{-8.1}$ &$35^{+16}_{-11}$ & $105^{+21}_{-20}$& $30.9^{+8.9}_{-8.4}$& $77^{+11}_{-10}$\\
		Ar & 4.193 & $4.8^{+3.4}_{-1.4}$  &$7.6^{+2.8}_{-2.8}$ &$5.7^{+4.9}_{-3.4}$ & $63^{+17}_{-34}$& $14^{+11}_{-11}$& $32^{+10}_{-10}$\\
		Ca & 3.359 & $9.6^{+3.0}_{-2.9}$  &$9.3^{+4.5}_{-4.0}$ &$9.9^{+5.9}_{-3.7}$ & $100^{+30}_{-28}$& $<26$& $64^{+17}_{-16}$\\
		Cr & 2.290 & $<2.8$  &$<8.9$ &$<6.5$ & $66^{+66}_{-64}$& $<72$& $<39$\\
		Mn & 2.102 & $6^{+12}_{-5}$ &$<14$ &$<3.2$ & $<132$& $<62$& $<70$\\
		Fe (K$\alpha$) & 1.936 & $178^{+17}_{-16}$  &$216^{+28}_{-25}$ &$160^{+28}_{-27}$ & $3570^{+150}_{-150}$& $939^{+77}_{-75}$& $1720^{+70}_{-70}$\\
		Fe (K$\beta$) & 1.757 & $23^{+39}_{-11}$ &$49^{+22}_{-21}$ &$<24$ & $1070^{+230}_{-200}$& $139^{+123}_{-63}$& $368^{+105}_{-120}$\\
		Ni & 1.658 & $14^{+13}_{-7}$  &$<21$ &$<27$ & $260^{+230}_{-150}$& $135^{+96}_{-91}$& $<89$\\
	\enddata
	
\end{deluxetable*}

\begin{deluxetable*}{c|c|c|c|c|c|c|c}
	\label{tab:gx_agn_flux}
	\tablecaption{Measured fluorescence line fluxes for GX 301-2, Mrk\,3, NGC 1068 and The Circinus Galaxy}
	\tablehead{\colhead{\textbf{Element}} & \colhead{\textbf{Rest Frame}} &\multicolumn{5}{c}{\textbf{Photon Flux ($10 ^{-6}$ ph s$^{-1}$ cm $^{-2}$)}}\\ \colhead{}& \textbf{Wavelength (\AA)} &\textbf{GX 301-2 IM}&\textbf{GX 301-2 NA}&\textbf{GX 301-2 PP}&\textbf{Circinus}&\textbf{Mrk\,3}&\textbf{NGC\,1068}}
	\startdata
	Mg &  9.890 & $1.7^{+2.4}_{-1.6}$& $2.7^{+2.3}_{-1.9}$& $<1.4$& $0.5^{+0.4}_{-0.3}$&$1.2^{+0.5}_{-0.4}$ &$ 3.0^{+1.0}_{-1.0}$\\
	Si & 7.126 & $8.1^{+1.9}_{-1.9}$ &$19.1^{+2.4}_{-2.6}$  &$9.1^{+2.5}_{-2.2}$ & $4.8^{+0.5}_{-0.5}$& $4.3^{+0.5}_{-0.5} $& $8.0^{+0.8}_{-0.9}$\\
	S & 5.373 &  $22^{+7}_{-11}$ &$34^{+7}_{-8}$ & $30.1^{+4.0}_{-6.0}$& $4.5^{+0.9}_{-0.8}$& $3.3^{+1.3}_{-1.3}$& $3.8^{+0.9}_{-1.0}$\\
	Ar & 4.193 &  $9.3^{+4.3}_{-3.7}$ &$28^{+10}_{-10}$ &$29.7^{+3.7}_{-7.1}$ & $3.3^{+0.6}_{-0.8}$& $1.1^{+0.8}_{-0.4}$& $1.8^{+0.7}_{-0.6}$\\
	Ca & 3.359 &  $18^{+6}_{-5}$&$67^{+23}_{-21}$ &$65^{+9}_{-9}$ & $5.8^{+0.5}_{-0.5}$& $1.7^{+0.8}_{-0.5}$& $1.5^{+0.9}_{-0.7}$\\
	Cr & 2.290 & $<15$ &$57^{+34}_{-35}$ &$150^{+32}_{-31}$ & $3.3^{+1.0}_{-1.0}$&  $<1.2$& $ 0.8^{+0.5}_{-0.8}$\\
	Mn & 2.102 & $23^{+26}_{-18}$&$<62$ & $<45$ & $2.1^{+0.9}_{-1.6}$& $2.0^{+1.3}_{-1.1}$& $0.9^{+0.9}_{-0.7}$\\
	Fe (K$\alpha$) & 1.936 &$883^{+59}_{-56}$  &$3680^{+100}_{-100}$ &$11470^{+270}_{-280}$ & $319.7^{+3.2}_{-9.9}$& $48.4^{+3.9}_{-3.8}$& $49.5^{+3.5}_{-3.2}$\\
	Fe (K$\beta$) & 1.757 & $199^{+81}_{-48}$ &$765^{+135}_{-93}$ &$3460^{+250}_{-210}$ & $44.1^{+4.4}_{-4.0}$& $8.6^{+3.2}_{-2.8}$& $5.3^{+1.9}_{-2.5}$\\
	Ni & 1.658 & $<64$  &$180^{+180}_{-130}$ &$730^{+170}_{-310}$ & $ 15.9^{+3.8}_{-3.4}$& $3.2^{+3.0}_{-2.0}$& $<3.6$\\
	\enddata

\end{deluxetable*}

\begin{deluxetable*}{c|c|c|c|c|c}
	\label{tab:4151_flux}
	\tablecaption{Measured fluorescence line fluxes for NGC\,4151}
	\tablehead{\colhead{\textbf{Element}} & \colhead{\textbf{Rest Frame}} &\multicolumn{3}{c}{\textbf{Photon Flux ($10 ^{-6}$ ph s$^{-1}$ cm $^{-2}$)}}\\ \colhead{}& \textbf{Wavelength (\AA)} &\textbf{03-2000}&\textbf{03-2007}&\textbf{02-2014}&\textbf{03-2014}}
	\startdata
	Mg & 9.890 & $<3.3$ &$<2.1$  &$2.8^{+1.2}_{-1.2}$ & $1.7^{+1.8}_{-1.2}$ \\
	Si & 7.126 & $8.0^{+2.1}_{-2.2}$ &$5.1^{+1.4}_{-1.9}$  &$7.4^{+1.2}_{-1.2}$ & $10.5^{+1.9}_{-2.1}$\\
	S & 5.373 & $5.0^{+7.0}_{-3.4}$  &$6.9^{+5.3}_{-2.5}$ &$7.7^{+2.3}_{-2.2}$ & $8.3^{+3.8}_{-3.2}$\\
	Ar & 4.193 & $<14$  &$6.8^{+3.8}_{-3.6}$ &$8.4^{+4.0}_{-3.4}$& $<4.6$\\
	Ca & 3.359 & $13^{+11}_{-9}$ &$5.6^{+4.4}_{-3.5}$ &$6.4^{+5.0}_{-4.0}$ & $18^{+11}_{-9}$\\
	Cr & 2.290 & $<490$ &$8.9^{+8.3}_{-7.6}$ &$<5.6$ & $<10.2$\\
	Mn & 2.102 & $<17$ &$<4.0$ &$<7.5$ & $<22.3$\\
	Fe (K$\alpha$) & 1.936 & $202^{+39}_{-28}$ &$97^{+21}_{-18}$ &$177^{+14}_{-14}$ & $152^{+24}_{-22}$\\
	Fe (K$\beta$) & 1.757 & $<37$ &$17^{+17}_{-14}$ &$32^{+13}_{-11}$ & $19^{+28}_{-12}$\\
	Ni & 1.658 & $29^{+32}_{-27}$  &$26^{+21}_{-17}$ &$12^{+11}_{-9}$ & $<31$\\
	\enddata

\end{deluxetable*}




\end{document}